\documentclass[prb,a4paper,showpacs,twocolumn]{revtex4}

\usepackage{amssymb}
\usepackage{amsmath}
\usepackage{amsfonts}
\usepackage{graphicx}
\usepackage{bm}
\usepackage{color}
\usepackage{multirow}
\usepackage{natbib}
\usepackage{hyperref}
\usepackage[normalem]{ulem}

\DeclareMathOperator{\tr}{tr}
\DeclareMathOperator{\Sp}{sp}

\begin{document}

\title{Transport in a two-dimensional disordered electron
liquid with isospin degrees of freedom}

\author{Igor S. Burmistrov}

\affiliation{L.D. Landau Institute for Theoretical Physics, Kosygina
  street 2, 117940 Moscow, Russia}

\affiliation{Condensed-matter Physics Laboratory, National Research University Higher School of Economics, 101000 Moscow, Russia}

\begin{abstract}
Recent theoretical results on transport in a two-dimensional disordered electron liquid with spin and isospin degrees of freedom are reviewed. A number of experimental features in temperature dependence of resistivity at low temperatures in Si-MOSFET, n-AlAs quantum well and double quantum well heterostructures is explained. Novel behavior of low-temperature resistivity at low temperature in electron systems with isospin degrees of freedom are predicted.
\end{abstract}

\maketitle


\section{Introduction}\label{Intro.Ch1.ISB}

Essence of phenomenon of Anderson localization is quantum interference that can fully suppress diffusion of a quantum particle in a random potential.\cite{Anderson1958ISB} Wave function of a particle with a given energy can be localized or delocalized in space. It leads to a possibility for existence of quantum phase transition with changing particle energy or parameters of the random potential. This quantum phase transition is usually termed as Anderson transition. Instead of quantum mechanical problem for a single particle one can consider a system of noninteracting electrons. Then the Anderson transition can occur under change of the chemical potential or dimensionless parameter  $k_F l$ where $k_F$ denotes the Fermi momentum and $l$ stands for the elastic mean free path. From the point of view of transport properties the phase with delocalized states is a metal whereas the phase with localized states is an insulator.

In general, three scenarios are possible: (i) the states at all energies are localized; (ii) the states at all energies are delocalized; (iii) the mobility edge between localized and delocalized states (Anderson transition) exists. In $d=1$ dimension all states are localized in any random potential, \cite{Mott1961ISB} (see also Ref.\cite{Froelich1985ISB}). In $d=3$ mobility edge and Anderson transition exists.\cite{Anderson1958ISB} Existence of Anderson transition in $d=2$ is more complicated question. Based on the relation of conductance to the response to change in the boundary conditions for a finite size system,\cite{Thouless1974ISB} the scaling theory for conductance has been constructed.\cite{Anderson1979ISB} This scaling theory was in agreement with direct diagrammatic calculations of conductance in weak disorder limit ($k_F l\gg 1$).\cite{Gorkov1979ISB,Abrahams1980ISB} The scaling theory of Ref.\cite{Anderson1979ISB} predicts that in $d=2$ all electronic states are localized. In $d>2$ Anderson transition occurs and conductance vanishes as a power law at the transition point.\cite{Wegner1976ISB} Scaling theory of Anderson transition allows to use for its description powerful tools developed for critical phenomena: 
low energy effective action and renormalization group (see e.g., Refs.\cite{AmitISB, ZinnJustinISB}). 

For the problem of Anderson localization low energy effective action is the so-called nonlinear $\sigma$-model (NL$\sigma$M).\cite{Wegner1979ISB,Schaefer1980ISB,Efetov1980ISB,Juengling1980ISB,McKane1981ISB,Efetov1982ISB} It describes diffusive motion of electrons on scales larger than the mean free path.\footnote{In $d=1$ for single channel case NL$\sigma$M approach is not applicable since localization length is given by the mean free path and there is no spatial scales at which diffusive motion exists.\cite{Berezinsky1973ISB}} The study of Anderson localization in the limit of weak disorder is much more convenient with the help of NL$\sigma$M  than using standard diagrammatic technique.\cite{Lee1985ISB,Efetov1983ISB} The former describes interaction of diffusive modes leading at the length scales larger than the elastic mean free path to logarithmic divergences in $d=2$.~\footnote{Diffusive modes include diffusons (two-particle propagator in particle-hole channel) and cooperons (two-particle propagator in particle-particle channel).}

Existence of Anderson transition depends not only on spatial dimensionality but also on symmetry of single-particle hamiltonian. In $d=2$ Anderson transition is absent for orthogonal Wigner-Dyson symmetry class,\cite{Wigner1951ISB,Dyson1962aISB,Dyson1962bISB} i.e. for hamiltonians that preserve time-reversal and spin-rotational symmetries. For example, in the case of symplectic symmetry class in which spin-rotational symmetry is broken Anderson transition occurs in $d=2$. In general, there are 10 different symmetry classes of single-particle hamiltonians describing (quasi)particle motion in a random potential.\cite{Zirnbauer1996ISB,Zirnbauer1997ISB,Heinzner2005ISB}    

For each symmetry class the corresponding NL$\sigma$M contains the topological terms in some spatial dimensions.\cite{Schnyder2008ISB,Schnyder2009ISB,Kitaev2009ISB} 
Their existence affects localization crucially. The text-book example is integer quantum Hall effect. Hamiltonian for a particle in $d=2$ in the presence of perpendicular magnetic field which breaks time-reversal symmetry corresponds to the  unitary class. Standard scaling theory in the unitary class predicts localization at large length scales.\cite{Brezin1980ISB}
Existence of the topological term\cite{Pruisken1983ISB,Pruisken1984ISB} in NL$\sigma$M for the unitary class in $d=2$ results in delocalized states and integer quantized Hall conductance.\cite{Pruisken1987ISB}

We emphasize that different aspects of Anderson localization discussed above exist in the problem of noninteracting electrons.\footnote{We refer a reader to Refs.\cite{Mirlin2000ISB,Mirlin2008ISB,Anderson50yearsISB} for a detailed exposition of current progress in studies of Anderson localization.} To realize a system of noninteracting electrons in a laboratory is not an easy task since one needs to exclude electron-electron and electron-phonon interactions.\footnote{Recently, experiments on Anderson localization in cold atoms in random optical potential have been performed (see, e.g. Refs.\cite{AspectISB,DelandeISB}).} At low temperatures electron-electron interaction becomes important. Inelastic processes due to electron-electron scattering with small (compared to temperature) energy transfer lead to destruction of phase coherence on time scales longer than the phase breaking time $\tau_{\phi}$.\cite{Thouless1977ISB,Anderson1979bISB,Altshuler1982ISB}\footnote{We note that in general phase breaking rate is different from the rate of electron-electron collisions (out-scattering rate in the kinetic equation approach
\cite{Schmid1974ISB,Altshuler1979ISB}) (see e.g. Ref.\cite{Blanter1996ISB}).} As temperature is lowered the phase breaking time  $\tau_{\phi}$  and corresponding spatial scale $L_\phi$ increase and diverge at zero temperature. At finite temperatures such that $L_\phi<L$, the phase breaking length plays a role of an effective system size. It yields the temperature dependence of conductance.\cite{Altshuler1982ISB}

In addition to influence on conductance through the phase breaking time, 
\footnote{Provided in the noninteracting approximation all single-particle states are localized, electron-electron interaction can lead to the finite-temperature transition between the phase with zero (low-temperature phase) and finite (high-temperature phase) conductivity.\cite{Fleshman1980ISB,Gornyi2005ISB,Basko2006ISB}
} electron-electron interaction results in strong temperature dependence of conductance at low temperatures due to virtual processes in electron-electron scattering.\cite{Altshuler1979cISB} Physically, strong temperature dependence of conductance is due to coherent scattering electrons off the Fridel oscillations.\cite{Zala2001ISB}. Strong (as compared with Fermi liquid) temperature dependence appears also in thermodynamic quantities, e.g. the specific heat and the static spin susceptibility (see e.g., Ref.\cite{Aronov-AltshulerISB}). The most interesting case is the case of  $d=2$ in which both weak localization and electron-electron\cite{Altshuler1980ISB} contributions to conductance are logarithmic in temperature. In the orthogonal symmetry class, contribution to the conductance due to strong electron-electron interaction is of opposite sign with respect to weak localization contribution. Therefore, in the presence of  \emph{electron-electron interaction} the metal-insulator quantum phase transition is possible in $d=2$.

The first attempt to  extend the scaling theory of Ref.\cite{Anderson1979ISB} to the metal-insulator transition in the presence of eletcron-electron interaction was performed in Ref.\cite{McMillan1981ISB}. In spite of confusion between the local density of states and the thermodynamic one, important outcome of Ref.\cite{McMillan1981ISB} was an idea of two-parameter scaling description for metal-insulator transition in the presence of interaction. The 
breakthrough was done in Ref.\cite{Finkelstein1ISB} where NL$\sigma$M was derived for the case of interacting electrons starting from the microscopic theory. With the help of renormalization group analysis of this, so-called  Finkelstein NL$\sigma$M the scaling theory of the metal-insulator transition in the presence of electron-electron interaction for $d>2$ was built.\cite{Finkelstein2ISB,Finkelstein3ISB,Finkelstein4ISB,Castellani1984ISB,Castellani1984bISB,Finkelstein5ISB} As usual, strong electron-electron interaction results in a change of the universality class of the metal-insulator transition in comparison with the noninteracting case (see e.g., Ref.\cite{FinkelsteinReviewISB,KirkpatrickBelitzISB}). In other words, electron-electron interaction is usually a relevant perturbation. We mention that recently the influence of electron-electron interaction on localization has been studied for the superconducting and chiral symmetry classes.\cite{DellAnna2006ISB}

In $d=2$ for the orthogonal symmetry class in the range of weak disorder the contribution to the conductance due to electron-electron interaction is stronger than weak localization contribution. In total, this leads to metallic behavior of conductivity at low temperatures.\cite{Finkelstein5ISB} This fact is in favor of existence of metal-insulator transition in $d=2$ in the presence of electron-electron interaction.

\begin{figure}[t]
\centerline{\includegraphics[width=0.45 \textwidth]{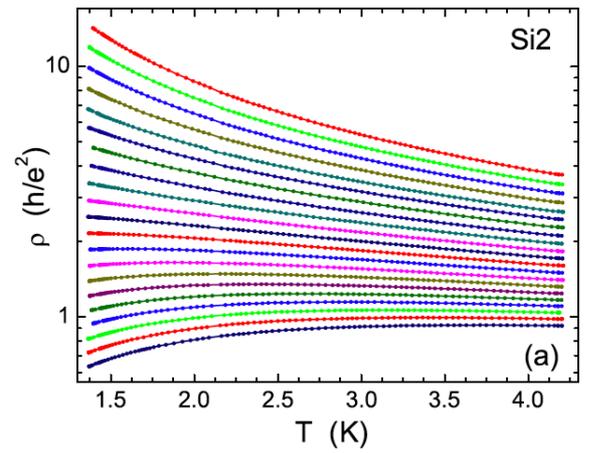}}
\caption{Temperature dependence of resistivity in Si-MOSFET. Electron concentration increases from top to bottom on $0.224\cdot 10^{10}$ cm$^{-2}$, starting from $6.72\cdot 10^{10}$ cm$^{-2}$ for the upper curve. Adapted from Ref.\protect\cite{PudalovPRLISB}.}
\label{Figure:Ch1:1:ISB}
\end{figure}

In spite of long history for experimental research of 2D electron systems,\cite{AFSISB} experimental observation of change in resistivity from insulating to metallic behavior with increase of electron density in Si metal-oxide-semiconductor field effect transistor (Si-MOSFET) (see Fig. \ref{Figure:Ch1:1:ISB}) became a surprise.\cite{Pudalov1ISB,prb95ISB}  
The observed temperature and electron density dependence of resistivity is as expected for resistivity in the vicinity of the metal-insulator transition. Later similar behavior of resistivity was experimentally observed in a variety of 2D electron systems.\cite{Abrahams2001ISB,Dolgopolov2003ISB,Kravchenko2004ISB,Pudalov2004ISB,Shashkin2005ISB,Gantmakher2008ISB} For the observed in Ref.\cite{Pudalov1ISB,prb95ISB} temperature behavior of resistivity an important  role seems to be played by the electron spin. Indication for this conclusion is the following experimental fact for Si-MOSFET: a weak (as compared with the field necessary for the full polarization) parallel magnetic field  changes behavior of the resistivity from metallic to insulating.\cite{Pudalov2ISB,VitkalovISB,Pudalov2003ISB} 

Similar effect of parallel magnetic field was observed experimentally in 2D hole system in GaAs/AlGaAs heterostructure.\cite{Yoon2000ISB} In the latter case, such behavior of resistivity with parallel magnetic field can be explained by the scaling theory which take into account that due to Zeeman splitting two among three triplet diffusive modes become massive.\cite{FinkelsteinReviewISB}. Recently, existence of the metal-insulator transition was demonstrated theoretically in $d=2$ for the special case in which the electron spin is equal to $\mathcal{N}\to \infty$ (the action of NL$\sigma$M is invariant under $SU(\mathcal{N})$ rotations).\cite{LargeNISB} It should be contrasted with the case of spinless electrons with Coulomb interaction for which the metal-insulator transition is absent in $d=2$.\cite{Baranov2002ISB} Therefore, spin degrees of freedom can crucially affect the low temperature behavior of resistivity and existence of the metal-insulator transition.

Recently, detailed comparison of two-parameter scaling theory of Ref.\cite{FinkelsteinReviewISB} with experimental data in Si-MOSFET has been performed. Temperature dependence of resistivity in metallic region,\cite{FPISB} 
magnetoresistance in parallel field,\cite{JETPL2ISB, KravchenkoNewISB} and resistance at criticality\cite{PudalovPRLISB} demonstrate reasonable agreement with the theory. We emphasize that 2D electron system in Si-MOSFET is specific since electrons occupy two valleys.\footnote{In 3D Si the  Brillouin zone contains 6 valleys which are degenerate in energy. In Si(001)-MOSFET degeneracy is lifted and only two valleys with centers at $z$ axis in the reciprocal space remains at low energies. The other four valleys are separated by the gap of the order of $200$ K and, therefore, do not participate in low temperature transport.} This leads to appearance of two additional energy scales: valley splitting 
$\Delta_v$ and inter valley elastic scattering rate  $1/\tau_v$. In vicinity of the critical region they were estimated as  $\Delta_v\approx 1$ K and $1/\tau_v\approx 0.1$ K.\cite{KuntsevichISB,KlimovISB} As a consequence the theory which takes into account finite values of $\Delta_v$ and $1/\tau_v$ should be developed for detailed comparison with experimental data on transport below $1$ K in two valley electron system such as Si-MOSFET. 

We mention that recently 2D two-valley electron system in SiO$_2$/Si(100)/SiO$_2$ quantum well has been experimentally studied.\cite{Renard2011ISB,Renard2013ISB} However, in Ref.\cite{Renard2013ISB} the inter valley scattering rate was estimated to be of the order of $7$ K such that 2D electron system behaves as single valley one for transport at low temperatures ($T\lesssim 7$ K).

It is well-known\cite{Shayegan0ISB} that two-valley electron system can be also realized in n-AlAs quantum well.\footnote{In 3D AlAs there are 6 valley which are degenerate in energy. Their centers are exactly at the boundaries of the Brillouin zone such that there are 3 valleys per each zone. In n-AlAs quantum well degeneracy is partially lifted and and only the valleys with centers at $x$ and $y$ axes in the reciprocal space contribute to low temperature transport. Therefore, there are effectively two valleys per Brllouin zone that participate in low temperature transport in $n-AlAs$ quantum well.}
Contrary to Si-MOSFET, in n-AlAs quantum well one can affect not only spin degrees of freedom by applying parallel magnetic field but isospin degrees of freedom (due to two valleys) also by means of stress that controls valley splitting.\cite{Shayegan1ISB,Shayegan2ISB} In Refs.\cite{Shayegan1ISB,Shayegan2ISB} it was found that in order to change temperature dependence of resistivity from metallic to insulating it is not enough to apply either parallel magnetic field or stress only. We mention that this experimental fact is in agreement with the experiments in Si-MOSFET.\cite{Pudalov2ISB,VitkalovISB,Pudalov2003ISB} Indeed in Si-MOSFET finite valley splitting is built in. However, this experimental results contradict to the theory of Ref.\cite{JETPL1ISB} with straight forward 
generalization of the theory of Ref.\cite{FPISB} to the case of finite spin or valley splitting only. According to results of Ref.\cite{JETPL1ISB} spin or valley splitting only is enough to change behavior of the resistivity from  metallic to insulating. The theoretical results presented in this paper resolve this discrepancy between theory and experiment.

One more system of 2D electrons which has isospin degrees of freedom is double quantum well heterostructure. In this case isospin distinguishes states localized in different quantum wells. In spite of large amount of interesting phenomena in double quantum well heterostructures such as Coulomb drag,\cite{DragExp1ISB,DragExp2ISB,DragExp3ISB,DragExp4ISB} exciton condensation,\cite{Current1ISB,Current2ISB,Current3ISB} ferromagnetic\cite{Fer1ISB,Fer2ISB} and counted antiferromagnetic\cite{AFerISB} phases, low temperature electron transport in the vicinity of possible metal-insulator transition has not been studied experimentally in details. Electron transport was studied only at large electron concentrations in the metallic region.\cite{MinkovOldISB,PortalISB} 

In double quantum well heterostructure with common scatterers for electrons one can expect behavior of resistivity similar to one in Si-MOSFET. Control of electron concentration in one quantum well by gate voltage allows to 
interpolate between the case of equal electron concentrations in each quantum well (akin to two valley system) and the case in which one quantum well is empty (akin to single valley system). According to the theory of Ref.\cite{FPISB}, 
one can expect strong change in temperature dependence of resistivity in the case of transfer from single to two valley situation. However, recent experiments\cite{MinkovGermanenko1ISB,MinkovGermanenko2ISB} in double quantum well heterostructures do not demonstrate significant difference in temperature dependence of resistivity during 
exhaustion of one of quantum wells. The theory presented in this paper allows us to explain this unexpected behavior of the resistivity measured in Refs.\cite{MinkovGermanenko1ISB,MinkovGermanenko2ISB}.

The paper is organized as follows. In Sec. \ref{Ch1:NLSM:ISB} we remind NL$\sigma$M approach and derive general RG equations in the one loop approximation in the case when an electron is characterized by both spin and isospin. In Sec. \ref{Si.Ch1:ISB} we use our general results for description of low temperature transport in 2D two valley electron system in Si-MOSFET in the presence of spin and valley splittings. In Sec. \ref{DQW.Ch1:ISB}, using our general results we explain transport experiments in 2D electron system in double quantum well heterostructure with common scatterers. We end the paper with Conclusions (Sec. \ref{Conc.Ch1:ISB}).

\section{Nonlinear $\sigma$-model\label{Ch1:NLSM:ISB}}

\subsection{Nonlinear $\sigma$-model action}

At low temperatures $T\tau_\textrm{tr}\ll 1$, where $\tau_\textrm{tr}$ denotes transport elastic scattering time,
nonlinear $\sigma$-model is the convenient tool for description of disordered electron liquid. NL$\sigma$M is designed to follow the interaction of low-energy ($|E| \lesssim 1/\tau_{\rm tr}$) bosonic modes termed as diffusons and cooperons. NL$\sigma$M is the field theory of the matrix field $Q^{\alpha\beta}_{mn}(\mathbf{r})$ which satisfies the following constraints: $Q^2(\mathbf{r})=1$, $\tr Q(\mathbf{r})=0$, 
$Q(\mathbf{r})=Q^\dag(\mathbf{r})$. The Greek indices $\alpha, \beta=1,2,\dots,N_r$ stand for the replica indices whereas the Latin indices are
integers $m, n$ corresponding to Matsubara energies $\varepsilon_n=\pi T (2n+1)$. In general, the matrix element 
$Q^{\alpha\beta}_{mn}(\mathbf{r})$ can have a matrix structure and satisfies additional constraints. Symbol $\tr$ denotes the trace over replica, Matsubara, spin and isospin indices. We refer a reader to Refs.\cite{FinkelsteinReviewISB,KirkpatrickBelitzISB} for a review on NL$\sigma$M approach. In what follows we assume the presence of a weak perpendicular magnetic field $B \gtrsim \max\{1/\tau_\phi,T\}/D$ where $D$ denotes the diffusion coefficient (see e.g., Ref.\cite{Aronov-AltshulerISB}) such that cooperons and interaction in the Cooper channel are suppressed. In this case $Q^{\alpha\beta}_{mn}(\mathbf{r})$ is a $4\times 4$ matrix in spin and isospin spaces.

The action of NL$\sigma$M is as follows
\begin{equation}
\mathcal{S} = \mathcal{S}_\sigma +\mathcal{S}_F+
\mathcal{S}_{SB} .
\label{Ch1:Sstart:ISB}
\end{equation}
Here the first term of the r.h.s. ($\mathcal{S}_\sigma$) describes NL$\sigma$M for the noninteracting electrons\cite{Wegner1979ISB,Schaefer1980ISB,Efetov1980ISB,McKane1981ISB}:
\begin{equation}
\mathcal{S}_\sigma = -\frac{\sigma_{xx}}{32} \int d\mathbf{r}\tr (\nabla Q)^2 ,
\label{Ch1:SstartSigma:ISB}
\end{equation}
where $\sigma_{xx}=4\pi\nu_\star D$ stands for the Drude conductivity in units $e^2/h$. The thermodynamic density 
of states $\nu_\star=m_\star/\pi$ is determined by the effective mass $m_\star$ (it includes Fermi-liquid corrections).
We mention that the dimensionless Drude conductivity is given by $\sigma_{xx}= 2 k_F l$, and it is assumed that $\sigma_{xx}\gg 1$. 

The presence of electron-electron interaction yields additional contribution to the NL$\sigma$M action\cite{Finkelstein1ISB,Finkelstein2ISB,Finkelstein3ISB,Finkelstein4ISB,BaranovISB}:
\begin{gather}
\mathcal{S}_F = - \pi T \int d \mathbf{r} \Bigl  [ \sum_{\alpha n; a b} \frac{\Gamma_{ab}}{4} \tr
I_n^\alpha t_{ab} Q(\mathbf{r}) \tr I_{-n}^\alpha t_{ab} Q(\mathbf{r})\notag \\
-4 z  \tr \eta Q + 6 z   \tr\eta\Lambda \Bigr ] .
 \label{Ch1:SstartF:ISB}
\end{gather}
Here $16$ matrices $t_{ab} = {\tau}_a \otimes {\sigma}_b$
($a,b=0,1,2,3$) are the generators of the $SU(4)$ group. The Pauli matrices ${\tau}_a$, $a=0,1,2,3$, acts on the isospin indices but the Pauli matrices ${\sigma}_b$, $b=0,1,2,3$, operates in the spin space. The quantities ${\Gamma}_{ab}$ stand for the interaction amplitudes. The parameter $z$ is independent charge (in the field theory sense) of NL$\sigma$M. It was introduced originally by Finkelstein in Ref.\cite{Finkelstein1ISB} in order RG flow to be consistent with the particle-number conservation. Its bare value is determined by the thermodynamic density of states: $z= \pi \nu_\star/4$. Physically, $z$ 
governs the temperature dependence of the specific heat\cite{CasDiCasISB}. The interaction amplitudes ${\Gamma}_{ab}$ in the action \eqref{Ch1:Sstart:ISB} are related with the Fermi-liquid interaction parameters ${F}_{ab}$: ${\Gamma}_{ab} = -z {F}_{ab}/(1+{F}_{ab})$. Matrices $\Lambda$, $\eta$ and $I_k^\gamma$ are defined as follows:
\begin{gather}
\Lambda^{\alpha\beta}_{nm} =
\mathrm{sign}\,n
\delta_{nm}\delta^{\alpha\beta} t_{00},\qquad
\eta^{\alpha\beta}_{nm} = n
\delta_{nm}\delta^{\alpha\beta}t_{00},\notag \\
(I_k^\gamma)^{\alpha\beta}_{nm} =
\delta_{n-m,k}\delta^{\alpha\gamma}\delta^{\beta\gamma}t_{00}.\label{Ch1:matrices_def:ISB}
\end{gather}

The last term in the r.h.s. of Eq.~\eqref{Ch1:Sstart:ISB} describes the effect of symmetry-breaking terms:
\begin{equation}
\mathcal{S}_{SB}=  i z_{ab} \Delta_{ab} \int
d\mathbf{r} \tr t_{ab} Q + \frac{N_r z_{ab} }{\pi T} \int
d\mathbf{r}\, \Delta_{ab}^2 . \label{Ch1:Sstartsb:ISB}
\end{equation}
For example, the Zeeman splitting $\Delta_s$ yields the term \eqref{Ch1:Sstartsb:ISB} with $t_{ab} = t_{03}$ and $\Delta_{03}=\Delta_s$. In the case of the two-valley 2D electron system in Si-MOSFET, the valley splitting $\Delta_{v}$ results in $\Delta_{30}=\Delta_{v}$ in Eq. \eqref{Ch1:Sstartsb:ISB}.

Matrix $Q$ in action~\eqref{Ch1:Sstart:ISB} is formally of the infinite size in the Matsubara space. However, in order to handle it one needs to introduce cut-off at large Matsubara frequencies. We assume that integers $m,n$ which correspond to Matsubara energies are restricted to the range $-N_M\leqslant m,n \leqslant N_M-1$ with $N_M\gg 1$. The condition of applicability of NL$\sigma$M gives the following estimate $N_M \sim 1/(2\pi T\tau_{\rm tr})$.

It is well-known~\cite{BaranovISB,KamenevAndreevISB} that global rotations of  $Q$ by matrix $\exp(i \hat\chi)$:
\begin{equation}
Q(\mathbf{r}) \to \exp(i \hat\chi) Q(\mathbf{r}) \exp(-i \hat\chi) ,\,\, \hat \chi = \sum_{\alpha,n} \chi^\alpha_n I^\alpha_n \label{Ch1:GenFRot:ISB}
\end{equation}
are important due to their relation with spatially constant electric potential. The latter can be gauged away by suitable gauge transformation of electron operators. Therefore it is convenient to define the limit $N_M\to \infty$ in a such way that the following relations hold ($\mathcal{F}$-algebra)~\cite{BaranovISB}:\footnote{We mention that the limit $N_M\to \infty$ should be taken at fixed $N=N_r N_M$. Then, the latter vanishes in the replica limit $N_r \to 0$. NL$\sigma$M with a finite value of $N_M$ is equivalent to NL$\sigma$M for noninteracting electrons at length scale exceeding $\sqrt{D/2\pi T N_M}$ ~\cite{Pisma2005ISB}.}
\begin{eqnarray}
\tr I^\alpha_n  t_{ab} e^{i\hat\chi} Q e^{-i\hat\chi} &=& \tr I^\alpha_n  t_{ab} e^{i\chi_0} Q e^{-i\chi_0}
+ 8 i n (\chi_{ab})^\alpha_{-n}\,,\notag\\
\tr \eta e^{i\hat\chi} Q e^{-i\hat\chi} &=& \tr \eta Q +
\sum_{\alpha n;ab } i n
(\chi_{ab})^\alpha_n\tr I^\alpha_n  t_{ab}
Q \notag \\
&-&  4 \sum_{\alpha n;ab}
n^2
(\chi_{ab})^\alpha_n(\chi_{ab})^\alpha_{-n} . \label{Ch1:Falg:ISB}
\end{eqnarray}
Here $\chi_0 = \sum_{\alpha} \chi_0^\alpha I_0^\alpha$. With the help of relations~\eqref{Ch1:Falg:ISB} one can check that provided $\Gamma_{00}=-z$ the NL$\sigma$M action $\mathcal{S}_\sigma+\mathcal{S}_F$ is invariant 
under global rotations \eqref{Ch1:GenFRot:ISB} with $\chi_{ab} = \chi \delta_{a0}\delta_{b0}$.

\subsection{Physical observables}

The most important quantities that contain information on low-energy behavior of NL$\sigma$M are physical observables $\sigma_{xx}^\prime$, $z^\prime$, and $z_{ab}^\prime$. They correspond to the parameters $\sigma_{xx}$, $z$, and $z_{ab}$ of action~\eqref{Ch1:Sstart:ISB}. Here $\sigma_{xx}^\prime$ is the conductivity of electron system defined via linear response to an external electromagnetic field. The observable $z^\prime$ determines the 
specific heat\cite{CasDiCasISB} and $z_{ab}^\prime$ corresponds to the static generalized susceptibility,\cite{Finkelstein4ISB,Castellani1984bISB} $\chi_{ab} = 2 z_{ab}^\prime/\pi$, which are responses to $\Delta_{ab}$.

The conductivity $\sigma^\prime_{xx}$ is given by the Kubo formula\cite{BPSISB}
\begin{gather}
\sigma^\prime_{xx}(i\omega_n) = -\frac{\sigma_{xx}}{16 n}\left
\langle\tr[I_{n}^\alpha, Q][I_{-n}^\alpha, Q] \right
\rangle + \frac{\sigma_{xx}^2}{64 d n } \int
d\mathbf{r}^\prime \notag \\ \times\langle\langle \tr I_n^\alpha
Q(\mathbf{r})\nabla Q(\mathbf{r})\cdot \tr I_{-n}^\alpha
Q(\mathbf{r}^\prime)\nabla Q(\mathbf{r}^\prime)\rangle \rangle .
\label{Ch1:SigmaODef:ISB}
\end{gather}
Here the analytic continuation to the real frequencies, $i\omega_n \to \omega + i0^+$ and static limit $\omega \to 0$ are should be taken. The averages in Eq. \eqref{Ch1:SigmaODef:ISB} are with respect to the action~\eqref{Ch1:Sstart:ISB}, and $\langle\langle A\cdot B\rangle\rangle = \langle A B\rangle - \langle A\rangle \langle B\rangle$. The physical observable $z^\prime$ is determined by the thermodynamic potential $\Omega$~\cite{BPSISB}:
\begin{equation}
z^\prime = \frac{1}{2\pi \tr \eta \Lambda}\frac{\partial}{\partial
T}\frac{\Omega}{T}.\label{Ch1:Defz:ISB}
\end{equation}
The physical observables $z_{ab}^\prime$ can be found from the following relations:\cite{FinkelsteinReviewISB}
\begin{equation}
z^\prime_{ab} = \frac{\pi}{2 N_r} \frac{\partial^2
\Omega}{\partial \Delta_{ab}^2} .
\label{Ch1:Defzsv:ISB}
\end{equation}


\subsection{One loop renormalization}

\subsubsection{Perturbation theory}

In order to construct the perturbation theory in small parameter $1/\sigma_{xx}$ it is convenient to use the square-root parameterization of $Q$:
\begin{gather}
Q = W+\Lambda\sqrt{1-W^2},\qquad W =
  \begin{pmatrix}
    0 & w \\
    w^\dag & 0
  \end{pmatrix} .\label{Ch1:Qexp:ISB}
\end{gather}
Then action~\eqref{Ch1:Sstart:ISB} can be written as the infinite series in powers of $w$ and $w^\dag$ fields. In the absence of the symmetry-breaking term $\mathcal{S}_{SB}$ the propagator of fields $w$ and $w^\dag$ becomes:
\begin{gather}
\langle
[w_{ab}(\mathbf{q})]_{n_1n_2}^{\alpha_1\alpha_2}
[w_{cd}^\dag(-\mathbf{q})]_{n_4
n_3}^{\alpha_4\alpha_3}
\rangle
 = \frac{4}{\sigma_{xx}}
\delta_{ab;cd}
\delta^{\alpha_1\alpha_3}\delta^{\alpha_2\alpha_4} \notag \\
 \times
\delta_{n_{12},n_{34}}   D_q(\omega_{12})  \Bigl [
\delta_{n_1n_3}
-\frac{32 \pi  T \Gamma_{ab}}{\sigma_{xx}}
\delta^{\alpha_1\alpha_2} D^{(ab)}_q(\omega_{12}) \Bigr ] 
,\label{Ch1:Prop:ISB}
\end{gather}
%
where $\omega_{12}=\varepsilon_{n_1}-\varepsilon_{n_2} = 2\pi Tn_{12} = 2\pi T(n_1-n_2)$ and
\begin{gather}
D_q^{-1}(\omega_n) = q^2+\frac{16 z \omega_n}{\sigma_{xx}} , \notag \\
 [D^{(ab)}_q(\omega_n)]^{-1} = q^2 +\frac{16
(z+\Gamma_{ab})\omega_n}{\sigma_{xx}} \, .  
\label{Ch1:DDProp_def:ISB}
\end{gather}
Here and further on we use notations $n_1, n_3, \dots$ for non-negative integers and $n_2,
n_4, \dots$ for negative ones.

Dynamical generalized susceptibility $\chi_{ab}(\omega,\mathbf{q})$ describes linear response to the space and time dependent parameter $\Delta_{ab}$. It can be found from the following Matsubara susceptibility\cite{FinkelsteinReviewISB}
\begin{gather}
\chi_{ab}(i\omega_n,\mathbf{q}) =\frac{2 z_{ab}}{\pi} - T z_{ab}^2 \langle\langle \tr
I^\alpha_n t_{ab} Q(\mathbf{q}) \notag \\
\times \tr I^\alpha_{-n} t_{ab}
Q(-\mathbf{q})\rangle \rangle\label{Ch1:SS1:ISB}
\end{gather}
with the help of analytic continuation to real frequencies: $i\omega_n \to
\omega + i0^+$~\cite{Finkelstein4ISB}. In the tree level approximation Eq.~\eqref{Ch1:SS1:ISB} 
becomes
\begin{equation}
\chi_{ab}(i\omega_n,\mathbf{q}) = \frac{2 z_{ab}}{\pi} \left (1 -
\frac{16 z_{ab} \omega_n}{\sigma_{xx}} D^{ab}_q(\omega_n)\right ).\label{Ch1:ChisTreeLevel:ISB}
\end{equation}

In some cases of matrix $\Gamma_{ab}$ the action ÄÅÊÓÔ×ÉÅ $\mathcal{S}_\sigma+\mathcal{S}_F$ can be invariant under global rotations $Q \to u Q u^{-1}$ with a spatially independent matrix  
$u = u_{a_0b_0} t_{a_0b_0}$. Provided such invariance of the action exists, the quantity  
$\tr t_{a_0b_0} Q$ is conserved. As a consequence, the corresponding Ward identity holds: $\chi_{a_0b_0}(\omega,\mathbf{q}=0)=0$. Then Eq.~\eqref{Ch1:ChisTreeLevel:ISB} implies that
$z_{a_0b_0} = z+\Gamma_{a_0b_0}$. In general case, there is no simple relation between $z_{ab}$ and $\Gamma_{ab}$.

The symmetry-breaking term $\mathcal{S}_{SB}$ with $\Delta_{a_0b_0}$ changes Eq. \eqref{Ch1:Prop:ISB}. The propagator reads 
\begin{gather}
\langle
[w_{ab}(\mathbf{q})]_{n_1n_2}^{\alpha_1\alpha_2}
[w_{cd}^\dag(-\mathbf{q})]_{n_4
n_3}^{\alpha_4\alpha_3}
\rangle
 = \frac{4}{\sigma_{xx}}\delta^{\alpha_1\alpha_3}\delta^{\alpha_2\alpha_4}
\delta_{n_{12},n_{34}}
 \notag \\
 \times 
 \Biggl \{ \widehat D_q(\omega_{12}) \Bigl [
\delta_{n_1n_3}
-\frac{32 \pi  T}{\sigma_{xx}}
\delta^{\alpha_1\alpha_2}  \widehat\Gamma \widehat{D}^{(\rm int)}_q(\omega_{12}) \Bigl ] \Biggr \}_{ab,cd}
,\label{Ch1:Prop2:ISB}
\end{gather}
where $\widehat\Gamma = \textrm{diag}\, \{\Gamma_{00},\Gamma_{01},\Gamma_{02},\Gamma_{03},\Gamma_{10},\dots,\Gamma_{33}\}$ and
\begin{gather}
\bigl[\widehat D_q(\omega_{n})\bigr]^{-1}_{ab,cd} = D_q^{-1}(\omega_{n}) \delta_{ab,cd}+ \frac{8 i z_{a_0b_0} \Delta_{a_0b_0}}{\sigma_{xx}} \mathcal{C}_{ab,cd}^{a_0b_0}, \notag \\
\bigl[\widehat D^{({\rm int})}_q(\omega_{n})\bigr]^{-1}_{ab,cd} = 
\bigl[\widehat D_q(\omega_{n})\bigr]^{-1}_{ab,cd} + \frac{16
\Gamma_{ab}\omega_n}{\sigma_{xx}}\delta_{ab,cd} .
 \label{Ch1:DD_Dmod:ISB}
\end{gather}
Here $\mathcal{C}_{cd;ef}^{ab}$ denotes the $SU(4)$ structure constants:  $[t_{cd}, t_{ef}] = \sum_{ab} \mathcal{C}_{cd;ef}^{ab} t_{ab}$. šAs one can find from Eq. 
\eqref{Ch1:DD_Dmod:ISB}, a part of diffusive modes becomes massive and does not lead to logarithmic divergencies at length scales $L\gg \sqrt{\sigma_{xx}/(z_{a_0b_0}\Delta_{a_0b_0})}$. One can check that it is the modes determining renormalization of the corresponding generalized susceptibility $\chi_{a_0b_0}(\omega,\mathbf{q})$. šTherefore, at length scales $L\gg \sqrt{\sigma_{xx}/(z_{a_0b_0}\Delta_{a_0b_0})}$ the physical observable $z_{a_0b_0}^\prime$ does not acquire renormalization.

\subsubsection{One loop renormalization of physical observables}

At length scales $L\ll~\min\limits_{ab} \sqrt{\sigma_{xx}/(z_{ab}\Delta_{ab})}$ one can neglect the symmetry-breaking term $\mathcal{S}_{SB}$. Using Eq.~\eqref{Ch1:Prop:ISB} we compute averages in Eq.~\eqref{Ch1:SigmaODef:ISB} in the one loop approximation. After analytic continuation we obtain
\begin{align}
\sigma^\prime_{xx} & = \sigma_{xx} 
+\frac{64}{\sigma_{xx} d} \Im  \int \frac{d^d \mathbf{p}}{(2\pi)^d}\, p^2 \sum_{ab}  \Gamma_{ab}  
\int d\omega \notag \\ & \times  \frac{\partial}{\partial\omega} \Bigl ( \omega \coth \frac{\omega}{2T} \Bigr )
[D^R_p(\omega)]^2 
D^{(ab),R}_p(\omega) .
\label{Ch1:sigma22:ISB}
\end{align}
Here $D_p^R(\omega)$ and $D^{(ab),R}_p(\omega)$ denote retarded propagators corresponding to Matsubara propagators $D_p(\omega_n)$ and $D_p^{(ab)}(\omega_n)$, respectively:
\begin{gather}
[D_p^R(\omega)]^{-1}  = p^2- \frac{16 i\omega  z}{\sigma_{xx}} \,,  \notag \\
 [D^{(ab),R}_p(\omega)]^{-1} = p^2 -\frac{16i\omega 
(z+\Gamma_{ab})}{\sigma_{xx}} . \label{Ch1:CondFinDD}
\end{gather}

In order to find the physical observable $z^\prime$, one has to evaluate the thermodynamic potential $\Omega$. In the  one loop approximation we find
\begin{gather}
T^2\frac{\partial\Omega/T}{\partial T}=  8 N_r T\sum_{\omega_n>0}
\omega_n
 \Bigl [z+\frac{2}{\sigma_{xx}} \sum_{ab} \int  \frac{d^d \mathbf{p}}{(2\pi)^d} \notag \\ \times  \Bigl ( (z+\Gamma_{ab}) D_p^{(ab)}(\omega_n)-z D_p(\omega_n)
\Bigr )\Bigr
].\label{Ch1:z1:ISB}
\end{gather}
Hence, using Eq.~\eqref{Ch1:Defz:ISB}, we obtain
\begin{gather}
z^\prime =z +\frac{2}{\sigma_{xx}}\sum_{ab} \Gamma_{ab} \int  \frac{d^d \mathbf{p}}{(2\pi)^d}
D^R_p(0)  . \label{Ch1:z2:ISB}
\end{gather}
The physical observables $z_{ab}^\prime$ can be found from the generalized susceptibilities $\chi_{ab}(\omega,\mathbf{q})$  in the static limit
$\omega=0$ and $q\to 0$. In the one loop approximation we find
\begin{gather}
z^\prime_{ab} = z_{ab}  + \frac{ 32 \pi z_{ab}^2}{\sigma_{xx}^2} \sum_{cd;ef} \left [  \mathcal{C}_{cd;ef}^{ab} \right ]^2 
 \int   \frac{d^d \mathbf{p}}{(2\pi)^d} T \sum_{\omega_m>0} \notag \\ \times 
 \Bigl [ 
D^{(ef)}_p(\omega_m) D_p^{(cd)}(\omega_m) - D_p^2(\omega_m) \Bigr ] . \label{Ch1:zabR:ISB}
\end{gather}
In general, even in the absence of the symmetry-breaking term $\mathcal{S}_{SB}$, the observables $z_{ab}$, $z$ and $\Gamma_{ab}$ are unrelated. Therefore, renormalization of  $\Gamma_{ab}$ needs to be found by other means. For example, it can be done by the background field renormalization method\cite{AmitISB}. Then we obtain
\begin{gather}
\Gamma_{ab}^\prime = \Gamma_{ab}  - \int  \frac{d^d \mathbf{p}}{(2\pi)^d} D_p(0) \sum_{cd;ef}
\frac{\Gamma_{cd} }{8 \sigma_{xx}}\bigl [  \Sp (t_{cd} t_{ef} t_{ab}) \bigr ]^2 \notag \\
- \frac{32 \pi T}{\sigma_{xx}^2} \sum_{\omega_m>0} \int  \frac{d^d \mathbf{p}}{(2\pi)^d}
\sum_{cd;ef} \left [  \mathcal{C}_{cd;ef}^{ab} \right ]^2  \Bigl \{\Gamma_{ab}^2 D_p^2(\omega_m)  \notag \\
- \Bigl [ \Gamma_{cd}\Gamma_{ef}+\Gamma_{ab}^2-2\Gamma_{ab}\Gamma_{cd}\Bigl ]
 D_p^{(cd)}(\omega_m)D_p^{(ef)}(\omega_m)  \Bigr \} . \label{Ch1:RenGam1:ISB}
\end{gather}
Here $\Sp$ denotes trace over the spin and isospin spaces. 


%
%
\subsection{One loop RG equations\label{Sec1:2:4:ISB}}

As usual (see e.g., Ref.\cite{AmitISB}) we derive the following one loop renormalization group equations in $d=2$ from Eqs.~\eqref{Ch1:sigma22:ISB}-\eqref{Ch1:RenGam1:ISB}:\cite{Burmistrov2011bISB}
\begin{align}
\frac{d \sigma_{xx}}{d y} & =- \frac{2}{\pi} \left [ 2 + \sum_{ab} f(\Gamma_{ab}/z)\right ],  \quad y=\ln L/l \notag \\
\frac{d \Gamma_{ab}}{dy} &= -\frac{1}{2\pi\sigma_{xx}}
 \sum_{cd;ef} \Biggl [ 
 \bigl [  \Sp (t_{cd} t_{ef} t_{ab}) \bigr ]^2\frac{\Gamma_{cd}}{8} + \left [  \mathcal{C}_{cd;ef}^{ab} \right ]^2 \notag \\
 &\times  \Bigl (\frac{\Gamma_{ab}^2}{z} - \frac{(\Gamma_{ab}-\Gamma_{cd})(\Gamma_{ab}-\Gamma_{ef})}{\Gamma_{cd}-\Gamma_{ef}}\ln \frac{z+\Gamma_{cd}}{z+\Gamma_{ef}} \Bigr ) \Biggr] ,\notag \\
\frac{d z}{d y} &= \frac{1}{\pi\sigma_{xx}} \sum_{ab} \Gamma_{ab} , \quad f(x) = 1- \frac{1+x}{x}\ln(1+x). \label{Ch1:RG_1loop:ISB}
\end{align}
RG equations \eqref{Ch1:RG_1loop:ISB} describe behavior of the physical observables at length scales $l\ll L\ll \min\limits_{ab} \sqrt{\sigma_{xx}/(z_{ab}\Delta_{ab})}$ at $T=0$. They generalize previous results for two-valley electron system~\cite{FinkelsteinReviewISB,FPISB} to the case of  different interaction amplitudes $\Gamma_{ab}$. It is important to mention that the symmetric situation in which all $\Gamma_{ab}$ except $\Gamma_{00}$ are equal is unstable (see Appendix \ref{App:Ch1:ISB}). As we shall discuss further, different values of $\Gamma_{ab}$ can be realized experimentally 
in electron systems with spin and isospin degrees of freedom. RG equations \eqref{Ch1:RG_1loop:ISB} lead to a number of new effects as compared to the standard case:\cite{FinkelsteinReviewISB,FPISB} $\Gamma_{ab}=\Gamma$ for $(ab)\neq (00)$.

We note that we have added $2$ into the square brackets of the r.h.s. of RG equation for the conductivity. This term describes the weak-localization contribution due to cooperons. As it is well-known\cite{Aronov-AltshulerISB,AALK81ISB,AA81ISB}, this contribution is insensitive to symmetry-breaking terms $\mathcal{S}_{SB}$.\footnote{To be precise, it is true in the absence of spin (or isospin) flips which results in appearance in the action terms $\tr [\Sigma,Q]^2$ where matrix $\Sigma$ is determined by a particular (iso)spin  flip mechanism.} 
We remind that the weak-localization contribution is suppressed by a weak perpendicular magnetic field $B \gtrsim 1/D\tau_\phi$.  Also cooperons contribute to the conductivity renormalization due to interaction in the Cooper channel\cite{FinkelsteinReviewISB}. However, in 2D electron systems the Cooper channel interaction is repulsive and renormalizes to zero. In addition, weak perpendicular magnetic field $B \gtrsim T/D$ suppresses such contributions\cite{AKLLISB}.

Using Eq. \eqref{Ch1:zabR:ISB} we find one loop RG equations for the physical observables $z_{ab}$:\cite{Burmistrov2011bISB} 
\begin{gather}
\frac{d z_{ab}}{dy} = \frac{z_{ab}^2}{2\pi\sigma_{xx}} \sum_{cd;ef} \frac{\left [  \mathcal{C}_{cd;ef}^{ab} \right ]^2}{\Gamma_{cd}-\Gamma_{ef}} 
\Bigl [ \ln\frac{z+\Gamma_{cd}}{z+\Gamma_{ef}} -\frac{\Gamma_{cd}-\Gamma_{ef}}{z}\Bigr ] . \label{Ch1:RG_1loop_zab:ISB}
\end{gather}
As one can see from Eqs.~\eqref{Ch1:RG_1loop:ISB}, the relation $z_{ab} = z+\Gamma_{ab}$ does not satisfy Eqs.~\eqref{Ch1:RG_1loop_zab:ISB}, generally.

Renormalization group equations~\eqref{Ch1:RG_1loop:ISB} describe length scale dependence of the physical observables at $T=0$. At finite temperature RG equations have to be stopped at the temperature induced length scale$L_{\rm in}$. In the case $\sigma_{xx}\gg 1$ and finite temperature RG equations ~\eqref{Ch1:RG_1loop:ISB}
are valid upto the  length scale $L_T = \sqrt{\sigma_{xx}/(zT)}$. At $L_T\ll L \ll L_\phi = \sqrt{\sigma_{xx}\tau_\phi /z}$ the conductivity is changed due to weak localization contribution only.

\subsection{Conductivity corrections due to small symmetry breaking terms}

Symmetry breaking term \eqref{Ch1:Sstartsb:ISB} does not affect the renormalization group equations at scales $L \ll  
\sqrt{\sigma_{xx}/(z_{ab}\Delta_{ab})}$ (or at temperatures $T\gg \Delta_{ab}$). However, they still change the temperature behavior of the physical observables. In the presence of the splitting $\Delta_{ab}$ the one-loop correction to the conductivity becomes
\begin{gather}
\sigma^\prime_{xx}  = \sigma_{xx} 
+\frac{64}{\sigma_{xx} d} \int \frac{d^d \mathbf{p}}{(2\pi)^d}\, p^2    
\int d\omega \frac{\partial}{\partial\omega} \Bigl ( \omega \coth \frac{\omega}{2T} \Bigr ) \notag \\  
\times  
\Im \Sp \Bigl ( [\widehat{D}^R_p(\omega)]^2 
\widehat{\Gamma} \widehat{D}^{({\rm int}),R}_p(\omega) \Bigr ).
\label{Ch1:sigma22Field:ISB}
\end{gather}
Hence, we find the following result for the second (the lowest) order in $\Delta_{ab}$ correction to the conductivity:
\begin{align}
\delta\sigma^\prime_{xx} & = \frac{128 z_{ab}^2 \Delta_{ab}^2}{\sigma^3_{xx} d}\int d\omega \frac{\partial}{\partial\omega} \Bigl ( \omega \coth \frac{\omega}{2T} \Bigr ) \sum_{cd} \Gamma_{cd} \notag \\
& \times  \Sp [t_{cd},t_{ab}]^2 \Im
 \int \frac{d^d \mathbf{p}}{(2\pi)^d}\, p^2    \notag \\
 & \times
 \frac{\partial^2}{(\partial p^2)^2} \Bigl ( [D^R_p(\omega)]^2 
D^{(cd),R}_p(\omega) \Bigr ).
\end{align}
Integrating over momentum and frequency, we obtain
\begin{align}
\delta\sigma^\prime_{xx} = \frac{3\zeta(3)}{128 \pi^3} \sum_{cd}  \Sp [t_{cd},t_{ab}]^2 \frac{\gamma_{cd}}{1+\gamma_{cd}} \left ( \frac{z_{ab}\Delta_{ab}}{z T} \right )^2.
\label{ch1:corcond:ISB}
\end{align}
Equation \eqref{ch1:corcond:ISB} generalizes the results \cite{Aronov-AltshulerISB,Castellani1998ISB,Aleiner2001ISB} for the correction to the magnetoresistance in small parallel magnetic field to the case of arbitrary interaction amplitudes. The parameters $z_{ab}$ and $z$, as well as the interaction amplitudes $\gamma_{cd}$ should be taken at the length scale $L_T = \sqrt{\sigma_{xx}/z T}$.

\subsection{Dephasing time}

One of the important characteristics of interacting electron system is the dephasing time.\cite{Aronov-AltshulerISB} In particular, the temperature dependence of the dephasing time determines the $T$ dependence of the weak-localization contribution to the conductivity. Generalizing known results\cite{Schmid1974ISB,Altshuler1979ISB,NarozhnyISB} to
the case of electrons with spin and isospin degrees of freedom we find the total dephasing rate at $T\gg \Delta_{s,v}$:\cite{Burmistrov2011bISB} 
\begin{gather}
\frac{1}{\tau_\phi} = -\frac{4}{\sigma_{xx}}\int\limits_{\tau_\phi^{-1}} d\omega \int \frac{d^2 q}{(2\pi)^2} \frac{1}{\sinh{(\omega/T)}} \Re D_q^R(\omega)  \notag \\
\times
\Im \sum_{ab} \frac{\Gamma_{ab}}{z} D_q^{(ab),R}(\omega) [D_q^R(\omega)]^{-1} .
\label{Ch1:DQW:Int1:ISB}
\end{gather}
where 
\begin{equation}
\mathcal{U}^{(ab)}(q,\omega) = \frac{\Gamma_{ab}}{z} D_q^{(ab),R}(\omega) [D_q^R(\omega)]^{-1} .
\end{equation}
Integrating over momentum and frequency and then cutting off the logarithmic divergence in the infra-red by the dephasing time we obtain for $1/\tau_\phi \gg \Delta_{s,v}$
\begin{gather}
\frac{1}{\tau_\phi} = \frac{T}{2\sigma_{xx}} \left ( \sum_{ab} \frac{\gamma_{ab}^2}{2+\gamma_{ab}} \right )  \ln T \tau_\phi .
\label{Ch1:DQW:TauPhi2:ISB}
\end{gather}
We mention that in Eq. \eqref{Ch1:DQW:TauPhi2:ISB} interaction amplitudes $\gamma_{ab}$ and  conductivity $\sigma_{xx}$ corresponds to the length scale $L_T = \sqrt{\sigma_{xx}/z T}$. 


\section{Spin-valley interplay in 2D disordered electron liquid\label{Si.Ch1:ISB}}

\subsection{Introduction}

In this section we use general RG Eqs. \eqref{Ch1:RG_1loop:ISB} for description temperature dependence of resistance and  spin/valley susceptibilities in 2D electron system in Si-MOSFET. We assume that there is a parallel magnetic field $B$ producing Zeeman splitting $\Delta_s=g_L\mu_B B\ll \tau_\textrm{tr}^{-1}$. Here $g_L$
and $\mu_B$ denotes $g$-factor and the Bohr magneton, respectively. Also, we assume that a finite valley splitting $\Delta_v$ and inter valley scattering time $\tau_v$ exist. We consider the case $\tau_{so}^{-1},\tau_v^{-1}\ll \Delta_v\ll \tau_\textrm{tr}^{-1}$ where $\tau_{so}$ stands for the spin relaxation time due to spin-orbit coupling. Experiments of Refs.\cite{KuntsevichISB,KlimovISB} indicate that such assumptions are reasonable for 2D electrons in Si-MOSFET with electron concentrations near the metal-insulator transition.

\subsection{Microscopic hamiltonian\label{Ch1:Si:Sec:Formalism:ISB}}

In order to describe 2D two-valley disordered electron system realized in Si(001)-MOSFET, it is convenient to write the electron annihilation operator with spin projection $\sigma/2$ on $z$ axis  as follows\cite{IordanskyISB,NestoklonISB}:
\begin{equation}
 \psi_\sigma(\mathbf{R}) = \sum_{\tau=\pm}
\psi^{\sigma}_{\tau}(\mathbf{r}) \varphi(z) [e^{i z
Q/2}+\tau e^{-i z
Q/2}]/\sqrt{2} .\label{WF:ISB}
\end{equation}
Here $z$ axis is perpendicular to the (001) plane,  
$\mathbf{r}$ denotes 2D coordinate vector, and $\mathbf{R} =
\mathbf{r}+z \mathbf{e_z}$. Subscript $\tau=\pm 1$ enumerates valley such that $\psi^\sigma_\tau$ denotes the annihilation operator for electron with $z$ axis spin projection $\sigma/2$ and isospin projection
$\tau/2$. We choose the envelope function 
$\varphi(z)$ to be normalized. In what follows, we neglect overlap $\int dz \, \varphi^2(z) \sin(Qz)$. Vector
$\mathbf{Q}=(0,0,Q)$ corresponds to the shortest distance between valleys in the reciprocal space. Its length can be estimated as $Q\sim
a^{-1}_\textrm{lat}$ where $a_\textrm{lat}$ stands for the lattice constant.\cite{AFSISB}

The 2D two-valley electron system is described by the following grand partition function:
\begin{equation}
Z = \int \mathcal{D}[\bar{\psi},\psi] \exp{{S}[\bar{\psi},\psi]},
\label{ch1:Z:ISB}
\end{equation}
where imaginary-time action reads ($\beta=1/T$)
\begin{gather}
{S} = -\int\limits_0^{\beta} dt \Biggl
\{ \int d\mathbf{r} \bar{\psi}^{\sigma}_{\tau}(\mathbf{r},t)\left 
[\partial_t +\mathcal{H}_0 \right ]{\psi}^{\sigma}_{\tau}(\mathbf{r},t) 
\notag \\
-\mathcal{L}_\textrm{dis}-\mathcal{L}_\textrm{int}\Biggr
\}.\label{Ch1:Si:Hstart:ISB}
\end{gather}
Single-particle hamiltonian
\begin{gather}
\mathcal{H}_0 =
-\frac{\nabla^2}{2m_e}-\mu + \frac{\Delta_s}{2}\sigma    +
\frac{\Delta_v}{2} \tau \label{Ch1:Si:HStart0:ISB}
\end{gather}
describes 2D quasiparticle with mass $m_e$ in the presence of parallel magnetic field $B$ ($\Delta_s=g_L\mu_B B$) and valley splitting $\Delta_v$. Here $\mu$ denotes the chemical potential. Lagrangian
\begin{equation}
\mathcal{L}_\textrm{dis} = -\int
d\mathbf{r}\,\bar{\psi}^{\sigma}_{\tau_1}(\mathbf{r})
V_{\tau_1\tau_2}(\mathbf{r}) {\psi}^{\sigma}_{\tau_2}(\mathbf{r})
\label{ch1:Ldis:ISB}
\end{equation}
encodes scattering electrons off a random potential $V(\mathbf{R})$. Matrix elements of the random potential can be written as
\begin{gather}
V_{\tau_1\tau_2}(\mathbf{r}) = \frac{1}{2}\int dz
\,V(\mathbf{R}) \varphi^2(z) 
\Bigl [ 1 + \tau_1\tau_2 \notag \\
+\tau_1 e^{i z Q} +\tau_2 e^{-iz Q}\Bigr ] . \label{Ch1:Si:Veqr:ISB}
\end{gather}
In general, matrix elements $V_{\tau_1\tau_2}(\mathbf{r})$ produce not only intra valley scattering but also inter valley scattering. We assume that random potential $V(\mathbf{R})$ is gaussian with $\langle V(\mathbf{R})\rangle = 0$ and
\begin{equation}
\langle
V(\mathbf{R}_1)V(\mathbf{R}_2)\rangle =
W(|\mathbf{r}_1-\mathbf{r}_2|,|z_1-z_2|) .
\label{ch1:VVV:ISB}
\end{equation}
Here function $W$ decays at typical length scale $d_W$. As one can check, if 
\begin{equation}
Q^{-1}\ll
d_W, \,\left [\int\varphi^4(z) dz\right ]^{-1} \ll n_e^{-1/2},\label{Ch1:Si:Condition1:ISB}
\end{equation}
the inter valley scattering is negligible and 
\begin{gather}
\langle
V_{\tau_1\tau_2}(\mathbf{r}_1)V_{\tau_3\tau_4}(\mathbf{r}_2)\rangle
= \frac{1}{2\pi\nu_\star\tau_i}
\delta_{\tau_1\tau_2}\delta_{\tau_3\tau_4}\delta(\mathbf{r}_1-\mathbf{r}_2),\notag \\
\frac{1}{\tau_i}=2\pi \nu_\star \int d^2\mathbf{r}dz_1 dz_2\,
W(|\mathbf{r}|,|z_1-z_2|) \varphi^2(z_1)\varphi^2(z_2). 
\end{gather}

Under conditions~\eqref{Ch1:Si:Condition1:ISB}, Lagrangian describing interaction is invariant under global $SU(4)$
rotations of electron operators $\psi^\sigma_\tau$ in spin and isospin spaces:
\begin{gather}
\mathcal{L}_\textrm{int} =-\frac{1}{2} \int
d\mathbf{r}_1d\mathbf{r}_2\,\bar{\psi}^{\sigma_1}_{\tau_1}(\mathbf{r}_1)
{\psi}^{\sigma_1}_{\tau_1}(\mathbf{r}_1)U(|\mathbf{r}_1-\mathbf{r_2}|) \notag \\
\times
\bar{\psi}^{\sigma_2}_{\tau_2}(\mathbf{r}_2) {\psi}^{\sigma_2}_{\tau_2}(\mathbf{r}_2).
\end{gather}
Here $U(r)=e^2/\varepsilon r$ where $\varepsilon$ stands for the dielectric constant. Low-energy part of $\mathcal{L}_\textrm{int}$ can be written as (see e.g., Refs.\cite{FinkelsteinReviewISB,KirkpatrickBelitzISB,Zala2001ISB})
\begin{gather}
\mathcal{L}_\textrm{int} = \frac{1}{4\nu_\star} \int\limits_{q l\lesssim 1} \frac{d\mathbf{q}}{(2\pi)^2}
\sum_{a,b=0}^3  \mathbb{F}_{ab}(q) m^{ab}(\mathbf{q}) 
m^{ab}(-\mathbf{q}) ,\notag \\
m^{ab}(\mathbf{q}) =  \int \frac{d\mathbf{k}}{(2\pi)^2} \overline{\Psi}(\mathbf{k}+\mathbf{q}) t_{ab} \Psi(\mathbf{k}) 
\label{ch1:Lint_Low_ISB}
\end{gather}
with $\overline{\Psi} = \{\bar\psi^+_+,\bar\psi^-_+,\bar\psi^+_-,\bar\psi^-_-\}$,  $\Psi = \{\psi^+_+,\psi^-_+,\psi^+_-,\psi^-_-\}^T$, and 
\begin{equation}
\mathbb{F}(q) = \begin{pmatrix}
F_s & F_t &F_t&F_t \\
F_t & F_t &F_t&F_t \\
F_t & F_t &F_t &F_t \\
F_t & F_t &F_t &F_t 
\end{pmatrix} .\label{Ch1:Si:Fmatrix:ISB}
\end{equation}
Here the quantity $F_t$ is a standard Fermi-liquid interaction parameter in the triplet channel. In the random phase approximation (RPA) it can be estimated as
\begin{gather}
F_t =  - \frac{\nu_\star}{2} \langle U^{\rm scr}(0)\rangle_{FS},\qquad 
F_s = \nu_\star U(q) +F_t ,\notag \\
\langle U^{\rm scr}(0)\rangle_{FS} = \int\limits_0^{2\pi} \frac{d\theta}{2\pi} U^{\rm scr}(2k_F\sin(\theta/2),0) ,
 \label{Ch1:Si:FFs:ISB} 
\end{gather}
where dynamically screened Coulomb interaction is as follows
\begin{align}
U^{\rm scr}(q,\omega) &=\frac{U(q)}{1+U(q) \Pi(q,\omega)}, \notag \\
\Pi(q,\omega) &= \frac{\nu_\star  D q^2}{D q^2-i\omega} . \label{Ch1:Si:ScrU}
\end{align}
  We mention that $k_F$ denotes the Fermi momentum of electrons 
in a single valley. The quantity $F_s$ involves Coulomb interaction at small momentum. In the limit $q\to 0$, $F_s(q)\approx \varkappa/q \to \infty$ where $\varkappa = 2\pi e^2 \nu_\star/\varepsilon$ stands for the inverse static screening length. We remind known results for the interaction parameter $F_t$:\cite{Zala2001ISB}
\begin{gather}
F_t = -\int\limits_0^{2\pi} \frac{d\theta}{4\pi} \frac{\varkappa }{2k_F\sin(\theta/2)+\varkappa} 
= -\frac{1}{2\pi} \mathcal{G}_0(\varkappa/2k_F) ,\notag \\
\mathcal{G}_0(x) = \frac{x}{\sqrt{1-x^2}} \ln \frac{1+\sqrt{1-x^2}}{1-\sqrt{1-x^2}} .\label{Ch1:Ft0:ISB}
\end{gather}
Under condition $\varkappa/k_F\ll 1$ which justifies RPA we obtain $\mathcal{G}_0(x)\approx x \ln (2/x)$.

\subsection{$SU(4)$ symmetric case}

It is convenient to introduce interaction $\gamma_t = -F_t/(1+F_t)$. Then for all interaction amplitudes $\Gamma_{ab}$ except $\Gamma_{00}$ the following relations hold: $\Gamma_{ab}= z\gamma_t$ for $(ab)\neq (00)$,  and $\Gamma_{00}=-z$. Therefore, at high energies $|E| \sim 1/\tau_{\rm tr}$ and short length scales $L \sim l$ electron-electron interaction in Si-MOSFET does not discriminate inter and intra valley interactions.
The presence in hamiltonian \eqref{Ch1:Si:HStart0:ISB} spin $\Delta_s$ and valley $\Delta_v$ splittings leads to the symmetry-breaking terms \eqref{Ch1:Sstartsb:ISB} in action \eqref{Ch1:Sstart:ISB}. At short length scales $L\ll L_{s,v}=\sqrt{\sigma_{xx}/(16 z_{s,v}\Delta_{s,v})}$ where $z_{s,v}=z(1+\gamma_t)$ or, equivalently, at high temperatures $T\gg \Delta_{s,v}$, the $SU(4)$ symmetry-breaking term \eqref{Ch1:Sstartsb:ISB} is not important. 

Using general one-loop RG equations \eqref{Ch1:RG_1loop:ISB}, we obtain the well-known results for 2D two-valley electron system~\cite{FPISB}:
\begin{eqnarray}
\frac{d\sigma_{xx}}{d y} &=& -\frac{2}{\pi} \left
[2+1+15 f(\gamma_t) \right ] ,\label{Ch1:Si:RG0_1:ISB}\\ \frac{d\gamma_t}{dy} &=&
\frac{(1+\gamma_t)^2}{\pi\sigma_{xx}},\label{Ch1:Si:RG0_2:ISB}\\
\frac{d\ln z}{d y} &=&
\frac{15\gamma_t-1}{\pi\sigma_{xx}}. \label{Ch1:Si:RG0_3:ISB}
\end{eqnarray}
Solution of RG Eqs. \eqref{Ch1:Si:RG0_1:ISB}-\eqref{Ch1:Si:RG0_2:ISB} yields metallic-type dependence of resistance $\rho=1/\pi \sigma_{xx}$ at large $y$ ($\rho$ decreases with increase of $y$). We mention that dependence $\rho(y)$ has the maximum at value of $y_{\rm max}$ such that $\gamma_t(y_{\rm max})\approx 0.46$.\cite{FPISB} The interaction amplitude $\gamma_t$ increases monotonically with $y$.

\subsection{$SU(2)\times SU(2)$ case
\label{Sec:Intermed:ISB}}

Let us assume that $\Delta_s\gg\Delta_v$. Then it is possible to consider the intermediate length scales $L_s\ll L \ll L_v$ for which the symmetry-breaking term  \eqref{Ch1:Sstartsb:ISB} with $\Delta_{03}=\Delta_s$ becomes important. If one use decomposition $Q=\sum_{ab} t_{ab} Q_{ab}$ then Eq.~\eqref{Ch1:Prop2:ISB} yields that modes $Q_{ab}$ with $b=1,2$ become massive and do not lead to logarithmic divergencies at  $L_s\ll L \ll L_v$. Therefore, at such length scales the NL$\sigma$M action is given by Eqs. \eqref{Ch1:SstartSigma:ISB} and \eqref{Ch1:SstartF:ISB} with 
\begin{equation}
Q = \sum_{a=0}^3\sum_{b=0,3}t_{ab} Q_{ab}. 
\end{equation}
We mention that if one defines matrix fields $Q_{\pm} = (t_{00}\pm t_{03})Q/2$ then they will be decoupled in the absence of electron-electron interaction. 

In the presence of Zeeman splitting one can distinguish interaction between electrons with equal spin projections and with different ones. It leads to the following form of matrix $\mathbf{\Gamma}$:
\begin{equation}
\mathbf{\Gamma} = \begin{pmatrix}
\Gamma_s & 0 &0 &\tilde{\Gamma}_t \\
\Gamma_t & 0 &0 &\Gamma_t \\
\Gamma_t & 0 &0 &\Gamma_t \\
\Gamma_t & 0 &0 &\Gamma_t 
\end{pmatrix} .
\end{equation}
Here combination $(\Gamma_s\pm \tilde{\Gamma}_t)$ describes interaction of electrons with equal/opposite spin projections.

Invariance of the action \eqref{Ch1:SstartF:ISB} with respect to global rotation \eqref{Ch1:GenFRot:ISB} with $\chi_{ab} = \chi \delta_{a0}\delta_{b0}$ is guaranteed by the condition $\Gamma_s=-z$. Conservation of the total isospin yields $z_v =
z+\Gamma_t$. Conservation of $z$-component of the total spin ($S_z$) results in $z_s = z+\tilde{\Gamma}_t$. Since renormalization of the spin susceptibility, i.e. $z_s$, is only possible due to interaction of diffusive modes with $S_z=\pm 1$ which are massive, there is no logarithmic divergences in $z_s$ at $L_s\ll L \ll L_v$. Therefore, we obtain
\begin{gather}
\frac{dz_s}{dy}=\frac{d (\Gamma_{s}-\tilde{\Gamma}_t)}{dy} = 0 .\label{Ch1:Si:G+-cons:ISB}
\end{gather}
We emphasize that observables $z_v$ and $z_s$ which behave in the same way at small length scales $L\ll L_s \ll L_v$, have to flow differently at $L_s\ll L \ll L_v$. This is the reason for the appearance of interaction amplitude $\tilde{\Gamma}_t$ which has different RG behavior in comparison with $\Gamma_t$.

Using general results \eqref{Ch1:RG_1loop:ISB}, we find the one loop RG equations for the intermediate length scales  $L_s\ll L \ll L_v$ ($\tilde{\gamma}_t=\tilde{\Gamma}_t/z$):\cite{BurmistrovChtchelkatchev2008ISB}
\begin{align}
\frac{d \sigma_{xx}}{dy} &=- \frac{2}{\pi}\left [ 2+1+6
f(\gamma_{t}) + f(\tilde{\gamma}_{t}) \right ] , \label{Ch1:Si:RG1_1:ISB}\\
\frac{d\gamma_t}{dy} &=
\frac{1+\gamma_t}{\pi\sigma_{xx}}(1+2\gamma_t-\tilde{\gamma}_t)\label{Ch1:Si:RG1_2:ISB} ,\\
\frac{d\tilde{\gamma}_t}{dy} &=
\frac{1+\tilde{\gamma}_t}{\pi\sigma_{xx}}(1-6\gamma_t-\tilde{\gamma}_t) ,\label{Ch1:Si:RG1_3:ISB}\\
\frac{d\ln z}{dy} &= -\frac{1}{\pi\sigma_{xx}} \left
(1-6\gamma_t-\tilde{\gamma}_t\right ).\label{Ch1:Si:RG1_4:ISB}
\end{align}
Here $y=\ln L/l_s$ where $l_s$ is of the order of $L_s$.\footnote{In order to find exact relation between $l_s$ and $L_s$ one needs to solve a complicated crossover problem (see e.g., Ref.\cite{AmitISB}) for description of behavior$\sigma_{xx}, \gamma_t, \tilde{\gamma}_t$ and $z$ at length scales $L \sim L_s$.} In what follows we assume that $\tilde{\gamma}_t(0)=\gamma_t(0)$. 

\begin{figure}[t]
\centerline{\includegraphics[width=0.45 \textwidth]{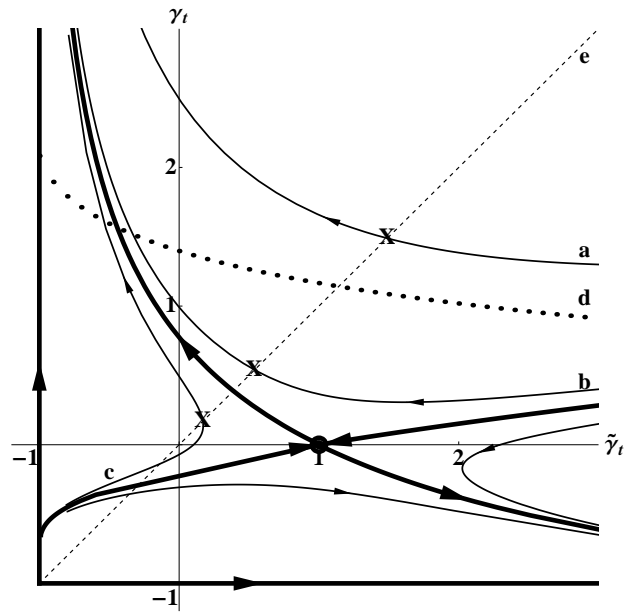}}  
\caption{Projection of three parameter $(\sigma_{xx},\tilde{\gamma}_t,\gamma_t)$ RG flow \eqref{Ch1:Si:RG1_1:ISB}-\eqref{Ch1:Si:RG1_3:ISB} to the plain $(\tilde{\gamma}_t,\gamma_t)$. Dotted curve $d$ is determined by equation $2+1+6f(\gamma_t)+f(\tilde{\gamma}_t)=0$. Dashed-line curve $e$ is described by equation $\gamma_t=\tilde{\gamma}_t$.}
 \label{Figure:Ch1:3:ISB}
\end{figure}

\begin{figure}[h]
 \centerline{\includegraphics[width=0.45 \textwidth]{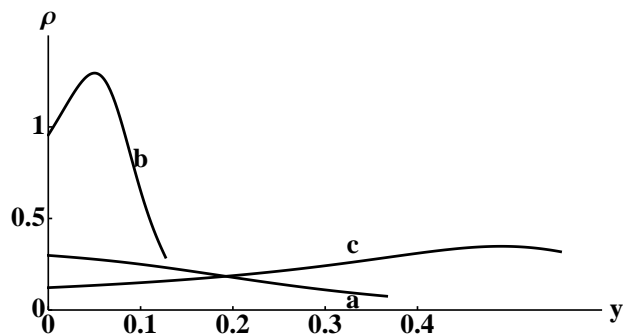}}
  \caption{Schematic dependence of resistivity $\rho=1/(\pi\sigma_{xx})$ on $y$. Initial values for the curves $a$, $b$, and $c$ correspond to crossing points $X$ of curves $a$, $b$, and $c$ with line $e$ in Fig.~\protect\ref{Figure:Ch1:3:ISB}. 
  }
  \label{Figure:Ch1:4:ISB}
\end{figure}

Figure~\ref{Figure:Ch1:3:ISB} illustrates RG flow in coordinates $(\tilde{\gamma}_t,\gamma_t)$. There is 
unstable fixed point at $\tilde{\gamma}_t=1$ and $\gamma_t=0$. However, in two-valley electron system this fixed point is unaccessible since RG flow starts near $\gamma_t=\tilde{\gamma}_t>0$. šAs shown in Fig.~\ref{Figure:Ch1:4:ISB}, two different regimes for behavior of the resistance $\rho$ are possible. Resistance decreases monotonously along the curve $a$ (see Fig.~\ref{Figure:Ch1:3:ISB}) which does not intersect the curve $d$ corresponding to equation $2+1+6f(\gamma_t)+f(\tilde{\gamma}_t)=0$. If one flows along the curves  $b$ and $c$ which intersect the curve $d$, then resistance has the maximum. We mention that resistance behavior at large values of $y$ is of metallic type. This is because the interaction amplitude  $\gamma_t$ grows at large $y$ whereas $\tilde{\gamma}_t$ tends to $-1$. Therefore, electrons with different spin projections become independent and Eqs.~\eqref{Ch1:Si:RG1_1:ISB}-\eqref{Ch1:Si:RG1_4:ISB} transforms into RG equations for two independent copies of single valley system. It is well-known\cite{FinkelsteinReviewISB} that in the case of single valley electron system RG equations yield metallic behavior of resistance at large values of $y$.

\subsection{Completely symmetry-broken case}

At large length scales $L\gg L_v\gg L_s$ the symmetry-breaking term \eqref{Ch1:Sstartsb:ISB} with $\Delta_{30}=\Delta_v$ becomes important also. We remind that there is eight massless modes ($Q_{ab}$ with $a=0,1,2,3$ and $b=0,3$) at intermediate length scales $L_s\ll L \ll L_v$. At large  length scales $L\gg L_v\gg L_s$ only four modes$Q_{00}, Q_{03}, Q_{30}$ and $Q_{33}$ remain massless. Therefore, NL$\sigma$M action at $L\gg L_v\gg L_s$ is given by Eqs. \eqref{Ch1:SstartSigma:ISB} and \eqref{Ch1:SstartF:ISB} with matrix
$Q = \sum_{a,b=0,3}t_{ab} Q_{ab}$. We mention that in the absence of electron-electron interaction four matrices 
$Q_{s}^{s^\prime} = (t_{00} +s t_{03}+s^\prime t_{30}+ss^\prime t_{33})Q/4$ where $s,s^\prime=\pm$ are decoupled. In the presence of strong spin and valley splitting one can distinguished interaction between electrons with equal and opposite spin and isospin projections. Hence, the matrix $\mathbf{\Gamma}$ acquires the following general form:
\begin{equation}
\mathbf{\Gamma} = \begin{pmatrix}
\Gamma_s & 0 &0 &\tilde{\Gamma}_t \\
0 & 0 &0 & 0  \\
0 & 0 &0 & 0 \\
\Gamma_t & 0 &0 &\hat{\Gamma}_t 
\end{pmatrix} .
\end{equation}
Here $(\Gamma_s+s \tilde{\Gamma}_t+s^\prime \Gamma_t+s s^\prime \hat{\Gamma}_t)/4$ corresponds to interaction of elections with spin projection $s$ and isospin projection $s^\prime$. Invariance of the NL$\sigma$M action under global rotation \eqref{Ch1:GenFRot:ISB} with $\chi_{ab} = \chi \delta_{a0}\delta_{b0}$ implies that the following condition holds: $\Gamma_s=-z$. Conservation of the $z$-th isospin component yields relation $z_v = z+\Gamma_t$, which holds also at the intermediate length scales $L_s\ll L \ll L_v$. Conservation of the $z$-th component of the spin results in the relation $z_s = z+\tilde{\Gamma}_t$. In addition to spin and valley susceptibility, one can define spin-valley susceptibility which describes the linear response to the $\Delta_{33}$ splitting. In the static limit it is determined by the quantity $z_{sv} = z+ \hat{\Gamma}_t$. At length scales $L\gg L_v\gg L_s$ spin, valley and spin-valley susceptibilities do not renormalize. Therefore, we obtain the following RG equations:
\begin{gather}
\frac{d z_v}{d y} = \frac{d (\Gamma_{t}-\Gamma_s)}{dy} = 0, \, 
\frac{dz_s}{dy} = \frac{d (\tilde{\Gamma}_t-\Gamma_{s})}{dy} = 0,\notag \\
\frac{dz_{sv}}{dy} =\frac{d (\hat{\Gamma}_t-\Gamma_{s})}{dy} =0.
\end{gather}
Since interaction amplitudes $\hat{\Gamma}_t$ and $\Gamma_t$ coincide at length scales $L\sim L_v$, and at larger scales they are renormalized in the same way, we consider them as equal:
$\hat{\Gamma}_t \equiv \Gamma_t$.

\begin{figure}[t]
\centerline{\includegraphics[width=0.45 \textwidth]{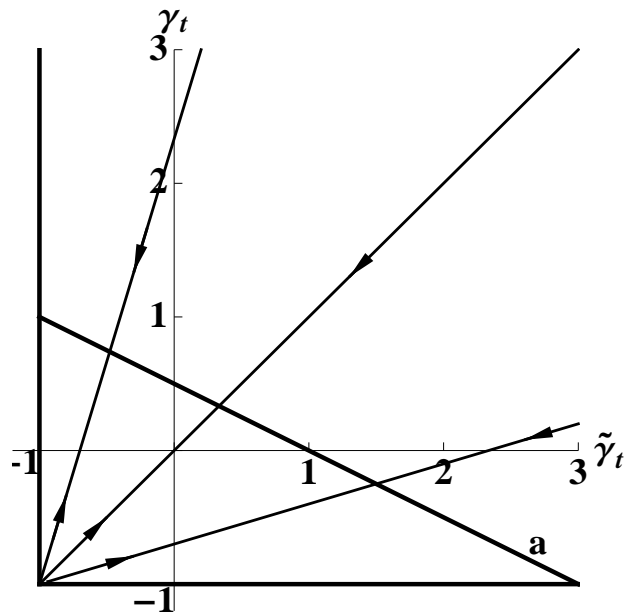}}  
\caption{Projection of three parameter $(\sigma_{xx},\tilde{\gamma}_t,\gamma_t)$ RG flow \eqref{Ch1:Si:RG3_1:ISB}-\eqref{Ch1:Si:RG3_3:ISB} to the plain
  $(\tilde{\gamma}_t,\gamma_t)$. Line $a$ corresponds to equation $2\gamma_t+\tilde{\gamma}_t=1$.
}
\label{Figure:Ch1:5:ISB}
\end{figure}

Using general RG Eqs. \eqref{Ch1:RG_1loop:ISB}, we obtain RG equation for length scales $L\gg L_v\gg L_s$:~\cite{BurmistrovChtchelkatchev2008ISB}
\begin{align}
\frac{d \sigma_{xx}}{d y} &=-
\frac{2}{\pi} \left [ 2+1+2 f(\gamma_{t}) +
f(\tilde{\gamma}_{t}) \right ], \label{Ch1:Si:RG3_1:ISB}\\
\frac{d\gamma_t}{d y} &=
\frac{1+\gamma_t}{\pi\sigma_{xx}}
(1-2\gamma_t-\tilde{\gamma}_{t}),\label{Ch1:Si:RG3_2:ISB}\\
\frac{d\tilde{\gamma}_{t}}{dy} &=
\frac{1+\tilde{\gamma}_{t}}{\pi\sigma_{xx}}(1-2\gamma_t-\tilde{\gamma}_{t}),\label{Ch1:Si:RG3_3:ISB}\\
\frac{d\ln z}{d y} &= -\frac{1}{\pi\sigma_{xx}} \left
[1-2\gamma_t-\tilde{\gamma}_{t}\right ] . \label{Ch1:Si:RG3_4:ISB}
\end{align}
Here $y=\ln L/l_v$ where the scale $l_v$ is of the order of $L_v$.  

RG flow for Eqs.~\eqref{Ch1:Si:RG3_1:ISB}-\eqref{Ch1:Si:RG3_3:ISB} in ($\gamma_t, \tilde{\gamma}_t$) plane is shown in Fig.~\ref{Figure:Ch1:4:ISB}. There is the line of fixed points determined by equation $2\gamma_t+\tilde{\gamma}_t=1$. RG flow in  ($\gamma_t, \tilde{\gamma}_t$) plane is just the straight  lines $(1+\tilde{\gamma}_t)/(1+\gamma_t)={\rm const}$. The curve described by equation $2+1+2f(\gamma_t)+f(\tilde{\gamma}_t)=0$ lies in the region of relatively large values of $\gamma_t$ and $\tilde{\gamma}_t$. Therefore, if initial values $\gamma_t(0)$ and $\tilde{\gamma}_t(0)$ are not large resistance $\rho(y)$ will be a monotonous increasing function of $y$, i.e. it will demonstrate insulating behavior.

\subsection{Discussion and comparison with experiments}

Keeping in mind discussion at the end of section \ref{Sec1:2:4:ISB} we assume that RG eqs.~\eqref{Ch1:Si:RG0_1:ISB}-\eqref{Ch1:Si:RG0_2:ISB}, \eqref{Ch1:Si:RG1_1:ISB}-\eqref{Ch1:Si:RG1_3:ISB} and \eqref{Ch1:Si:RG3_1:ISB}-\eqref{Ch1:Si:RG3_3:ISB} describe temperature dependence of the physical observables as listed in Table~\ref{Tab:Ch1:1:ISB}.

\begin{table}[t]
\caption{Temperature regions in which RG equations are applied.}
{\begin{tabular}{@{}ccc@{}} \toprule
RG Eqs. & $\Delta_s=0$& $\Delta_s\gg\Delta_v$ \\
\colrule
\eqref{Ch1:Si:RG0_1:ISB}-\eqref{Ch1:Si:RG0_2:ISB} &$\Delta_s\ll T\ll 1/\tau_{\rm tr}$  & $\Delta_s\ll T\ll 1/\tau_{\rm tr}$ \\
\eqref{Ch1:Si:RG1_1:ISB}-\eqref{Ch1:Si:RG1_3:ISB} & $1/\tau_v\ll T \ll \Delta_v$ & $\Delta_v \ll T \ll \Delta_s$ \\
\eqref{Ch1:Si:RG3_1:ISB}-\eqref{Ch1:Si:RG3_3:ISB} & & $1/\tau_v\ll T \ll \Delta_v$\\
\botrule
\end{tabular}}
\label{Tab:Ch1:1:ISB}
\end{table}

\begin{figure}[t]
\centerline{a) \includegraphics[width=0.45 \textwidth]{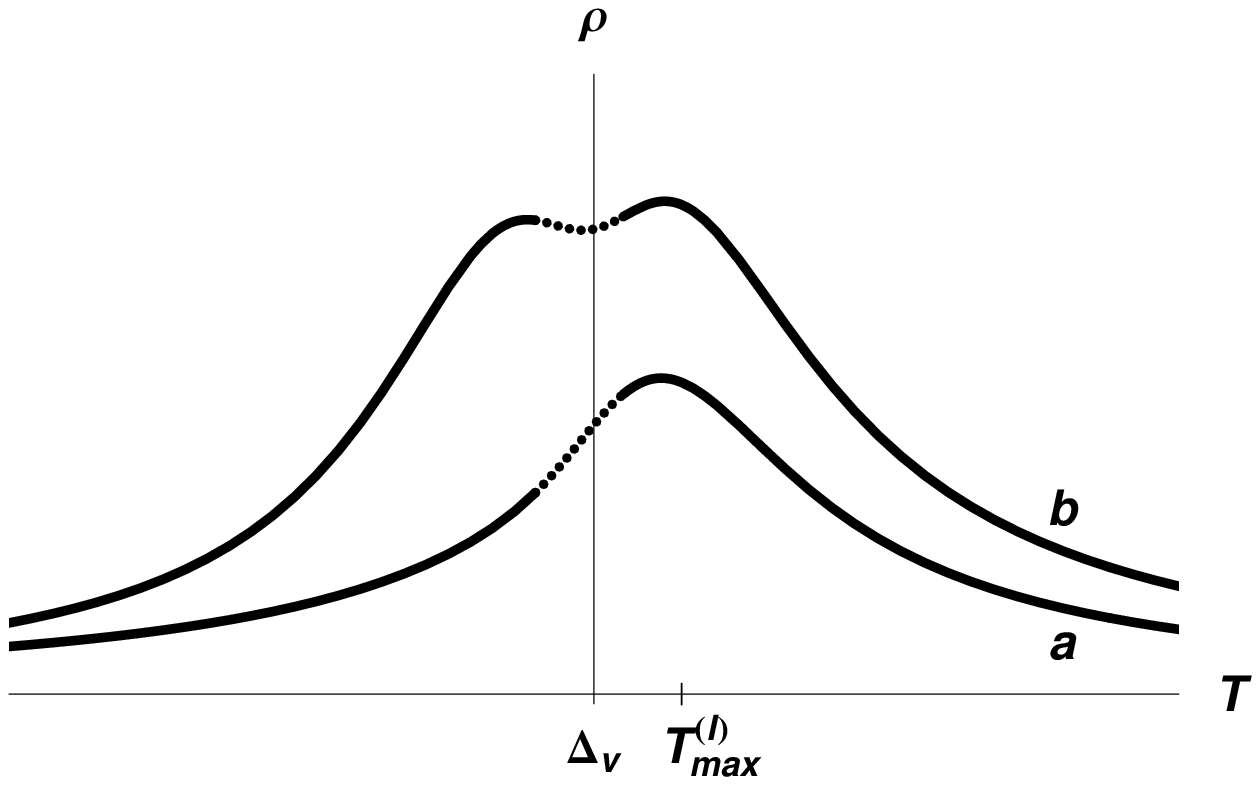}}
\centerline{b) \includegraphics[width=0.45 \textwidth]{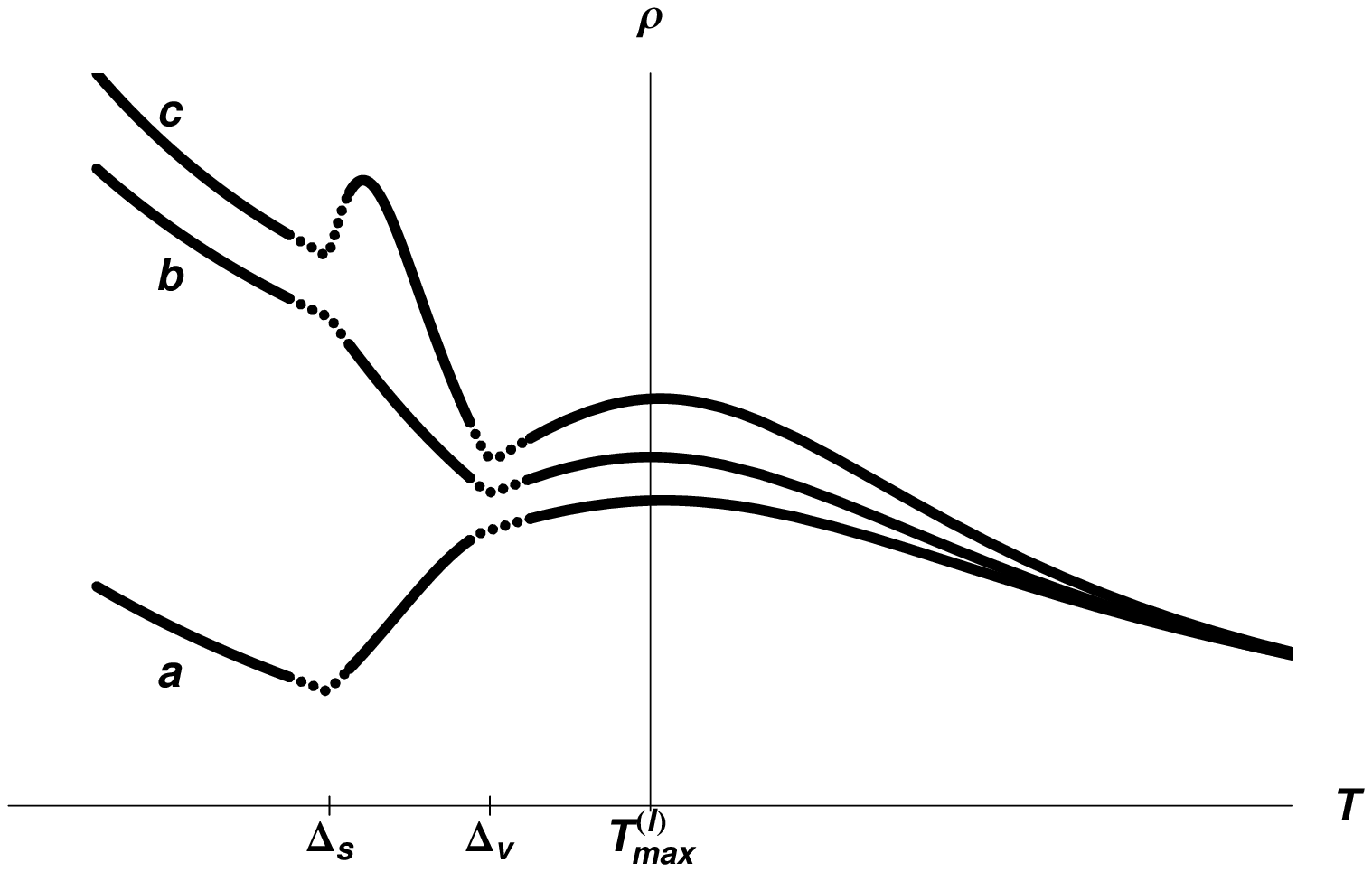}}
\centerline{c) \includegraphics[width=0.45 \textwidth]{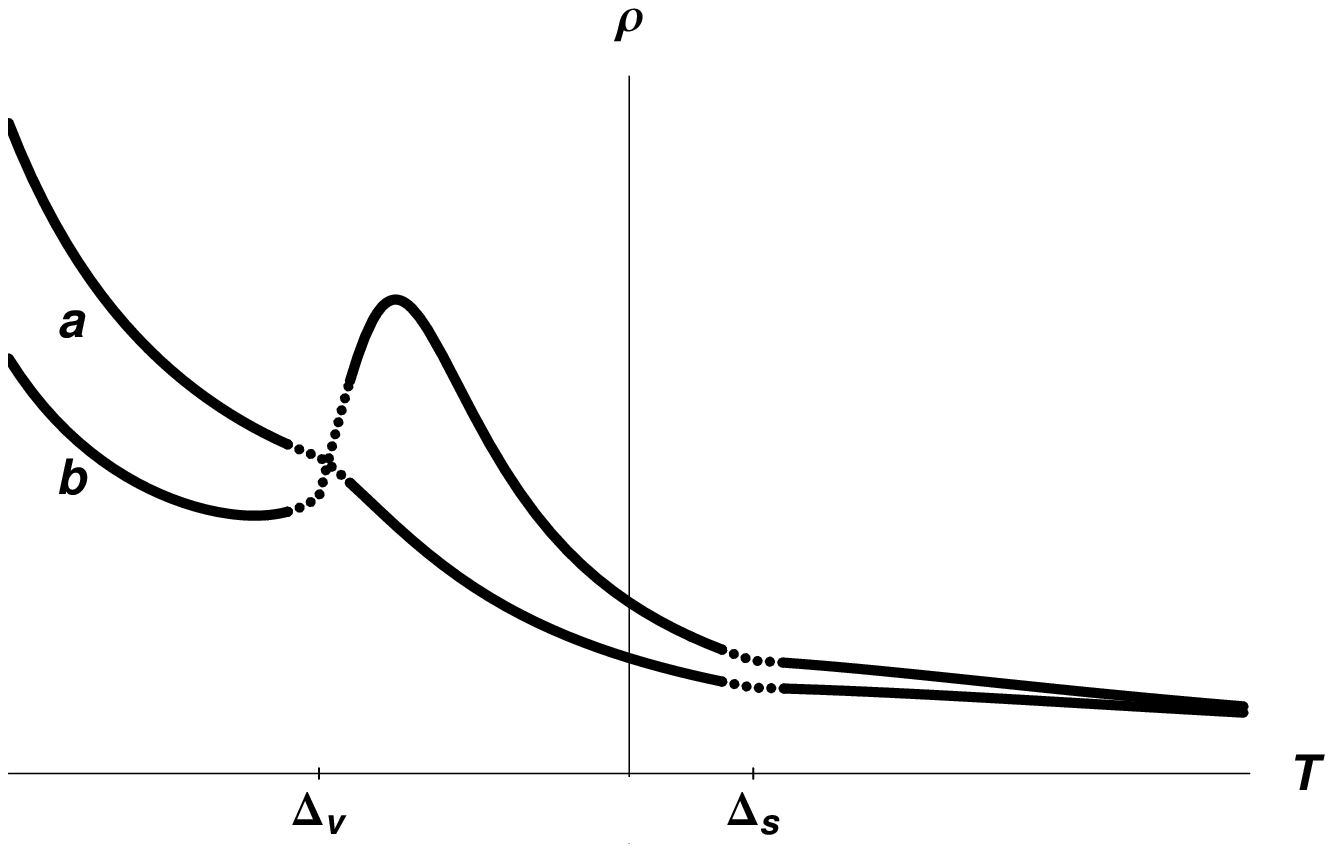}}
  \caption{Schematic dependence of resistivity $\rho(T)$ for a) zero spin splitting, b)  the case $\Delta_s<\Delta_v$, c) strong parallel magnetic field: $\Delta_v , T_\textrm{max}^{(I)}<\Delta_s$.}
  \label{Figure:Ch1:6:ISB}
\end{figure}

Let us start from the case of zero Zeeman splitting, $\Delta_s=0$. We assume that the following condition holds $\Delta_v < T_\textrm{max}^{(I)}$. Here $T_\textrm{max}^{(I)}$ denotes the temperature at which resistance reaches the maximum in accordance with Eqs. \eqref{Ch1:Si:RG0_1:ISB}-\eqref{Ch1:Si:RG0_2:ISB}.~\footnote{For example, in a Si-MOSFET sample with 2D electron concentration about  $10^{11}$\, cm$^{-2}$ the temperature $T_\textrm{max}^{(I)}$ is equal to a few Kelvins.} Then there are possible two type of behavior of resistance with temperature depending on the initial conditions (values of $\sigma_{xx}$ and $\gamma_t$ at temperature of the order of $1/\tau$). Both types of behavior are illustrated in Fig.~\ref{Figure:Ch1:6:ISB}a. šThe curve $a$ is typical for $\rho(T)$ dependence observed in experiments on Si-MOSFET.\cite{Pudalov1ISB} Remarkably there exists more complicated behavior: $\rho(T)$ has {\it two} maximums (the curve $b$ In Fig.~\ref{Figure:Ch1:6:ISB}a). We mention that such $\rho(T)$ dependence has not been observed experimentally yet. We emphasize that in the absence of Zeeman splitting metallic behavior of $\rho(T)$ at temperatures $1/\tau_v\ll T\ll 1/\tau$ does not spoiled by the presence of finite valley splitting $\Delta_v\gg 1/\tau_v$. This result is in agreement with experimental observations on 2D two-valley electron systems.\cite{Shayegan1ISB,Shayegan2ISB,Pudalov2ISB,VitkalovISB,Pudalov2003ISB}

In the presence of weak parallel magnetic field $\Delta_s < T_\textrm{max}^{(I)}$, three types of $\rho(T)$ behavior are possible as shown in Fig.~\ref{Figure:Ch1:6:ISB}b. In all three cases resistance has the maximum at temperature $T=T_\textrm{max}^{(I)}$ and increases as temperature decreases. As one can see in Fig.~\ref{Figure:Ch1:4:ISB}, for temperatures $\Delta_s\lesssim T\lesssim\Delta_v$ three different scenarios for $\rho(T)$  are possible: metallic (the curve $a$ in Fig.~\ref{Figure:Ch1:6:ISB}b), insulating (the curve $b$ in Fig.~\ref{Figure:Ch1:6:ISB}b) and nonmonotonous  (the curve $c$ in Fig.~\ref{Figure:Ch1:6:ISB}b). šIn the latter case, 
resistance can develop {\it two maximums} even in the presence of parallel magnetic field. At lower temperatures 
$1/\tau_v\ll T \lesssim \Delta_v$ resistance increases with decrease of $T$. Therefore, the presence of both valley and spin splittings ($\Delta_s<\Delta_v$) yields the change of $T$-dependence of resistance from metallic to insulating at low temperatures in agreement with experimental findings of Refs.\cite{Shayegan1ISB,Shayegan2ISB,Pudalov2ISB,VitkalovISB,Pudalov2003ISB}. We note that the typical temperature of metal-to-insulator crossover is of the order of either $\Delta_v$ or $\Delta_s$ depending on the initial values of $\sigma_{xx}$ and $\gamma_t$ at temperature of the order of $1/\tau_{\rm tr}$. 

In the presence of strong parallel magnetic field such that $\Delta_s > T_\textrm{max}^{(I)}$ the maximum in $\rho(T)$ is absent and only two types of resistance behavior occur as schematically illustrated in Fig.~\ref{Figure:Ch1:6:ISB}c. If the temperature $T_\textrm{max}^{(II)}$ of the resistance maximum found from RG Eqs. \eqref{Ch1:Si:RG1_1:ISB}-\eqref{Ch1:Si:RG1_3:ISB} is such that $T_\textrm{max}^{(II)}<\Delta_v$, then $\rho(T)$ is monotonously increasing with lowering temperature (the curve $a$ in Fig.~\ref{Figure:Ch1:6:ISB}c). In the opposite case $T_\textrm{max}^{(II)}>\Delta_v$, the $\rho(T)$ dependence is shown in Fig.~\ref{Figure:Ch1:6:ISB}c by the curve $b$. Therefore provided $\Delta_v > T_\textrm{max}^{(II)}$ the $\rho(T)$ dependence changes from metallic to insulating. It is in agreement with experimental data on magnetoresistance in Si-MOSFET.~\cite{Pudalov2ISB,JETPL2ISB} 
However, if valley splitting $\Delta_v < T_\textrm{max}^{(II)}$ then the maximum in $\rho(T)$ remains even in the presence of parallel magnetic field.

\begin{figure}[t]
\centerline{\includegraphics[width=0.45 \textwidth]{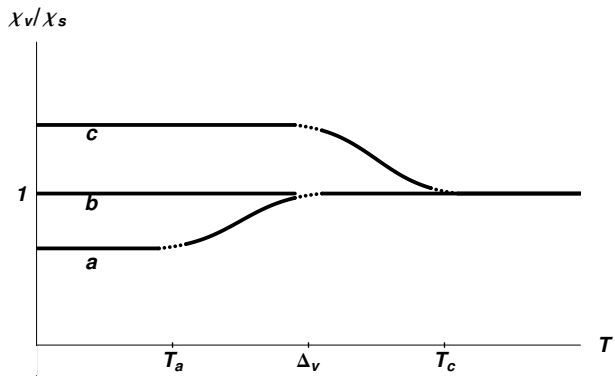}}  
\caption{Schematic dependence of the ratio $\chi_v/\chi_s$ on temperature in the cases i) $\Delta_s<\Delta_v$ (the curve $a$), ii) $\Delta_s=\Delta_v$ (the curve $b$), and iii) $\Delta_s>\Delta_v$ (the curve $c$). Characteristic temperature scales $T_{a,c} \equiv \Delta_s$.
}
\label{Figure:Ch1:9:ISB}
\end{figure}

In addition to interesting new temperature behavior of resistivity, RG equations derived above yield interesting predictions for $T$-dependence of the spin and valley susceptibilities. Let us consider the ratio of static valley and spin susceptibilities $\chi_v/\chi_s$. A schematic dependence of $\chi_v/\chi_s$ on $T$ is shown in Fig.~\ref{Figure:Ch1:9:ISB} at fixed values of valley splitting and different values of spin splitting. At high temperatures $T\gg \Delta_v, \Delta_s$ the ratio is equal to unity, $\chi_v/\chi_s=1$. At $T \ll \Delta_v, \Delta_s$, we obtain\begin{equation}
\frac{\chi_v}{\chi_s} 
  \begin{cases}
    < 1 & ,\, \Delta_s < \Delta_v ,\\
    =1 & ,  \,\Delta_s  = \Delta_v ,\\
    > 1  & ,\,\Delta_s > \Delta_v .
  \end{cases}
\end{equation}
Therefore, the ratio $\chi_v/\chi_s$ in the limit $T\to 0$ is sensitive to the ratio $\Delta_v/\Delta_s$.

In order to detect the second maximum in $\rho(T)$ experimentally one needs a system with wide interval between  $1/\tau_v$ and $\Delta_v$. Experimental data on magnetoresistance in Si-MOSFET~\cite{KuntsevichISB} allow to estimate inter valley scattering rate to be about $0.36$ K for electron concentrations in the range $3 - 6 \cdot 10^{11}$ cm$^{-2}$. In the same range of concentrations the valley splitting varies only weakly. However, there is significant sample-to-sample variation of the valley splitting.~\cite{KlimovISB} It was estimated to be in the range $0.4 - 0.7$ K. Due to such values of $1/\tau_v$ and $\Delta_v$ it is complicated to find the second maximum in $\rho(T)$ experimentally.~\cite{Punnoose1ISB}  We note that experiments in Si-MOSFET on magnetoresistance in parallel magnetic field demonstrates absence of dependence on relative orientation of current and magnetic field.\cite{Pudalov2003ISB} It signifies smallness of  effects due to spin-orbit coupling.

The influence of the spin and valley splittings on resistivity discussed above was due to change in RG equations corresponding to temperatures $T \ll \Delta_{s,v}$. In the opposite case of small spin and valley splittings, $\Delta_{s,v}\ll T$ there are corrections to resistivity proportional to $\Delta_s^2$ and $\Delta_v^2$ (see Eq. \eqref{ch1:corcond:ISB}).  
However, we mention that there is another source for the influence of $\Delta_{s,v}\ll T$ on temperature dependence of resistivity. It is due to possible dependence of initial (for RG flow) values of conductivity and interaction amplitudes on $\Delta_{s,v}$ which come from ballistic scales $T\gg 1/\tau_{\rm tr}$.\cite{JETPL2ISB} This mechanism accounts for the temperature dependence of magnetoresistivity in weak parallel magnetic field in experiments of Ref.\cite{JETPL2ISB}.

Finally, we reiterate the main result of the analysis presented in this section. The 2D two-valley disordered electron system realized in Si-MOSFET in the presence of spin and/or valley splittings has to be described by three-parameter RG equations at length scales $L\gg \min\{L_s,L_v\}$. Recently, RG Eqs. \eqref{Ch1:Si:RG1_1:ISB}-\eqref{Ch1:Si:RG1_3:ISB} and \eqref{Ch1:Si:RG3_1:ISB}-\eqref{Ch1:Si:RG3_3:ISB} were rederived in Refs.\cite{Punnoose2ISB,Punnoose3ISB}. Also in these papers the behavior of the resistance at $T \ll 1/\tau_v\ll\Delta_v$ has been investigated. It was found that in the absence of Zeeman splitting RG equations at $T \ll 1/\tau_v\ll\Delta_v$ are exactly the same as in the case of single valley system, i.e.
metallic behavior of $\rho(T)$ should persist down  to zero temperatures.\cite{FinkelsteinReviewISB} In the case of finite spin splitting $\Delta_s\gg\Delta_v\gg1/\tau_v$, at  $T \ll 1/\tau_v$ RG equations coincide with ones for single valley system in the presence of spin splitting $\Delta_s\gg T$. As well-known, under such circumstances the $\rho(T)$ dependence is of insulating type.\cite{FinkelsteinReviewISB}. šTherefore, one can conclude that as temperature crosses 
the scale $1/\tau_v$ the resistance does not change a type of its temperature dependence (metallic or insulating).

\section{2D disordered electron liquid in the double quantum well heterostructure \label{DQW.Ch1:ISB}}

\subsection{Introduction \label{Intro.DQW.Ch1:ISB}}

In this section we apply general RG Eqs. \eqref{Ch1:RG_1loop:ISB} to study of 2D electron system in a symmetric double quantum well with common disorder. We consider the case of equal electron concentrations and mobilities in both quantum wells. We shall assume that the following inequalities hold  $1/\tau_{+-}, \Delta_s,\Delta_{SAS} \ll T \ll 1/\tau_{\rm tr}$. Here $\Delta_{SAS}$ and $1/\tau_{+-}$ denote splitting and elastic scattering rate for transitions between symmetric and antisymmetric states in the double quantum well heterostructure, respectively. We note that such conditions correspond to the experiments of Refs.\cite{MinkovGermanenko1ISB,MinkovGermanenko2ISB}.


\subsection{Microscopic hamiltonian\label{Ch1:DQW:Sec_Form_Ham:ISB}}

In the case of symmetric double quantum well it is convenient to chose the following basis for electron annihilation operator:
\begin{equation}
\psi^{\sigma}(\mathbf{R}) = \psi_{\tau }^{\sigma}(\mathbf{r}) \varphi_\tau(z),\quad
\varphi_\tau(z)= \frac{\varphi_l(z)+\tau \varphi_r(z)}{\sqrt{2}} .\label{Ch1:DQW:WF:ISB}
\end{equation}
Here we assume that motion of electrons along $z$ axis is confined due to quantum well barriers. Vector $\mathbf{r}$ 
denotes coordinates in the plane perpendicular to the $z$ axis, and $\mathbf{R} = \mathbf{r}+z \mathbf{e_z}$. Superscript $\sigma=\pm$ stands for the spin projection to the $z$ axis, and subscript $\tau=\pm$ enumerates symmetric ($+$) and antisymmetric ($-$) states. Normalized wave functions
$\varphi_{l,r}(z)=\varphi(z\pm d/2)$ describe an electron localized in the left/right well. We neglect their overlap. Also we assume that both quantum wells are narrow such that  
\begin{equation}
\left [\int dz\, \varphi^4(z) \right ]^{-1} \ll d
\label{Ch1:DQW:Cond1:ISB}
\end{equation}
where $d$ is the distance between the centers of quantum wells.

In terms of electron operators $\psi_{\tau }^{\sigma}(\mathbf{r})$  and $\bar{\psi}_{\tau }^{\sigma}(\mathbf{r})$ the system is described by the grand partition function \eqref{ch1:Z:ISB} with the action  which has the same form as Eq. \eqref{Ch1:Si:Hstart:ISB}. The single particle hamiltonian is given by Eq. \eqref{Ch1:Si:HStart0:ISB} in which now $\Delta_{SAS}$ plays a role of $\Delta_v$. We mention that $\Delta_{SAS}$ can be estimated as $2 \varphi(d/2)\varphi^\prime(d/2)/m_e$~\cite{LL3ISB}. Scattering electrons off the random potential $V(\mathbf{R})$ is described by the Lagrangian $\mathcal{L}_\textrm{dis}$ given by Eq. \eqref{ch1:Ldis:ISB}. However, now the matrix elements $V_{\tau_1\tau_2}(\mathbf{r})$ are defined as
\begin{gather}
V_{\tau_1\tau_2}(\mathbf{r}) = \int dz\,
V(\mathbf{R}) \varphi_{\tau_1}(z)\varphi_{\tau_2}(z)
. \label{Ch1:DQW:Veqr:ISB}
\end{gather}
In general, matrix elements $V_{\tau_1\tau_2}$ produce transitions between symmetric and antisymmetric states in the double quantum well  heterostructure. We note that in the case of mirror symmetry: $V(\mathbf{r},z) = V(\mathbf{r},-z)$, the transitions between symmetric and antisymmetric states are absent. We assume that impurities that create the random potential $V(\mathbf{R})$ are situated in the plane $z=0$ which is in the middle between the quantum wells. This assumptions correspond to the double quantum well heterostructures studied in Refs.\cite{MinkovGermanenko1ISB,MinkovGermanenko2ISB}. Also we assume that $V(\mathbf{R})$ is gaussian with zero mean value and two-point correlation function given by Eq. \eqref{ch1:VVV:ISB}. Since for narrow quantum wells condition \eqref{Ch1:DQW:Cond1:ISB} holds, we can neglect the difference (of the order of $\varphi(d/2)\varphi^\prime(d/2)$) in scattering rates between only symmetric or only antisymmetric states. Then in the case of random potential with short-range correlations\footnote{In the experiments of Refs.\cite{MinkovGermanenko1ISB,MinkovGermanenko2ISB} random potential has been created by charged impurities situated near the plane $z=0$ (in the middle between the quantum wells). In this case the typical range for the random potential is of the order of 3D screening length, $d_W \sim 1/\sqrt{\varkappa k_F}$. Provided the condition $d_W\ll l$ (equivalently, $k_F l \gg \sqrt{k_F/\varkappa}$) holds we can consider the random potential as short-range correlated.} we find 
\begin{gather}
\langle
V_{\tau_1\tau_2}(\mathbf{r}_1)V_{\tau_3\tau_4}(\mathbf{r}_2)\rangle
= \frac{1}{2\pi\nu_\star\tau_i}
\delta_{\tau_1\tau_2}\delta_{\tau_3\tau_4}\delta(\mathbf{r}_1-\mathbf{r}_2),
\notag \\
\frac{1}{\tau_i}=2\pi \nu_\star \int d^2\mathbf{r}\,
W(|\mathbf{r}|,d/2,d/2) . \label{Ch1:DQW:RandPot1:ISB}
\end{gather}
Small asymmetry in impurity distribution with respect to the $z$ axis results in scattering between symmetric and antisymmetric states. Its rate can be estimated as $1/\tau_{+-} \sim b^2/(d^2\tau_i) \ll 1/\tau_i$ where $b$ denotes typical length scale characterizing  asymmetry. In what follows we assume that temperature $T\gg 1/\tau_{+-}$ and, therefore, we will neglect such scattering. 

The interacting Lagrangian is as follows
\begin{equation}
\mathcal{L}_{\rm int}=-\frac{1}{2}\int d \mathbf{R} d\mathbf{R^\prime} \rho(\mathbf{R}t)\, U(|\mathbf{R}-\mathbf{R^\prime}|)\, \rho(\mathbf{R^\prime}t)  ,
\label{ch1:Lint:gen:ISB}
\end{equation}
where $U(\mathbf{R}) = e^2/\varepsilon R$ and the electron density operator $\rho(\mathbf{R}t)=\bar{\psi}^{\sigma}_{\tau_{1}}(\mathbf{r}t)
\psi^{\sigma}_{\tau _{2}}(\mathbf{r}t)\varphi_{\tau _{1}}(z)\varphi _{\tau _{2}}(z)$. In the case of narrow quantum wells, Eq. \eqref{ch1:Lint:gen:ISB} becomes
\begin{gather}
\mathcal{L}_{\rm int}=-\frac{1}{8}\int d\mathbf{r} d\mathbf{r^\prime}\, 
\bar{\psi}^{\sigma_1}_{\tau_{1}}(\mathbf{r}t) \psi^{\sigma_1}_{\tau _{2}}(\mathbf{r}t)
\bar{\psi}^{\sigma_2}_{\tau_{3}}(\mathbf{r^\prime}t) \psi^{\sigma_2}_{\tau _{4}}(\mathbf{r^\prime}t)
\notag \\ \times 
\Bigl [ (1
+\tau_{1}\tau_{2}\tau_{3}\tau_{4}) U_{11}(|\mathbf{r}-\mathbf{r^\prime}|) +
\notag \\ 
+
(\tau_{1}\tau_{2}+\tau_{3}\tau_{4}) U_{12}(|\mathbf{r}-\mathbf{r^\prime}|) \Bigr] .
\end{gather}
Here
\begin{equation}
U_{11}(r) =\frac{e^2}{\varepsilon} \int dz dz^\prime \frac{\varphi_l^{2}(z) \varphi_l^{2}(z^\prime)}{\sqrt{r^2
+(z-z^\prime)^2}} \approx \frac{e^2}{\varepsilon r}
\end{equation}
describes electron-electron interaction inside a quantum well. Interaction of electrons from different quantum wells is given by
\begin{equation}
U_{12}(r) =\frac{e^2}{\varepsilon} \int dz dz^\prime \frac{\varphi_l^{2}(z) \varphi_r^{2}(z^\prime)}{\sqrt{r^2
+(z-z^\prime)^2}} \approx \frac{e^2}{\varepsilon \sqrt{r^2+d^2}} .
\end{equation}
Due to the difference between $U_{11}$ and $U_{12}$ Lagrangian $\mathcal{L}_{\rm int}$ is not invariant under global $SU(4)$ rotation of $\psi^\sigma_\tau$ in the spin and isospin spaces. It is the form of $\mathcal{L}_{\rm int}$  that makes the cases of electrons in the double quantum well heterostructure and Si-MOSFET to be different. The low-energy part of  $\mathcal{L}_\textrm{int}$ acquires the form of Eq. \eqref{ch1:Lint_Low_ISB} with 
the following 4$\times$ 4 matrix of the interaction parameters:
\begin{equation}
\bf{F}(q) = \begin{pmatrix}
F_s & F_t &F_t&F_t \\
\tilde F_s & F_t &F_t&F_t \\
F_v & F_v &F_v &F_v \\
F_v & F_v &F_v &F_v 
\end{pmatrix} .\label{Ch1:DQW:Fmatrix:ISB}
\end{equation}
In RPA the interaction parameters can be estimated as 
\begin{gather}
F_t  =  - \frac{\nu_\star}{2} \langle U_{11}^{\rm scr}(0)\rangle_{FS},\quad 
F_v =  -\frac{\nu_\star}{2} \langle U_{12}^{\rm scr}(0)\rangle_{FS} , \notag \\
F_s = \nu_\star [U_{11}(q) + U_{12}(q)] +F_t , \notag \\
\tilde{F}_s = \nu_\star [U_{11}(q) - U_{12}(q)] +F_t . \label{FFs:ISB} 
\end{gather}
Here $U_{11}(q) = 2\pi e^2/q\varepsilon$ and $U_{12}(q) = U_{11}(q) \exp(-qd)$. The quantities $F_t$ and  $F_v$ are analogous to the standard Fermi liquid parameter in the triplet channel. They are determined by the screened interaction 
 $U_{11/12}^{\rm scr}(q,\omega)$ averaged over the Fermi surface. In the case of equal electron concentrations and mobilities in both quantum wells, we find
 \begin{equation}
 \langle U_{11/12}^{\rm scr}(0)\rangle_{FS} = \int\limits_0^{2\pi} \frac{d\theta}{2\pi} U_{11/12}^{\rm scr}(2k_F\sin(\theta/2),0) ,
\end{equation}
where $k_F$ stands for the Fermi momentum in a quantum well. The interaction parameter $F_s$ involves Coulomb interaction at small transferred momentum. In the limit $q\to 0$, we obtain $F_s(q)\approx 2\varkappa/q \to \infty$ and $\tilde{F}_s = \varkappa d + F_t$. For $d=0$, the double quantum well transforms into the single quantum well with $\tilde{F}_s=F_t=F_v$. In this case, the electron system is equivalent to one in Si-MOSET. In the limit $d\to \infty$, intra well electron-electron interaction vanishes and $\tilde F_s=F_s$ but $F_v=0$. At finite value of $d$ and $\Delta_s=0$ the action becomes invariant under global $SU(2)$ spin rotations of electron operators in each well separately. 

Dynamically screened inter and intra well interaction in RPA are given as\cite{ZhengISB,KamenevOregISB,FlensbergISB}
\begin{gather}
U_{11}^{\rm scr} = \frac{U_{11} + \Pi_2[U_{11}^2 - U_{12}^2] }{1+[\Pi_1+\Pi_2]U_{11}+ \Pi_1\Pi_2[U_{11}^2 - U_{12}^2]} ,\notag
\\
U_{12}^{\rm scr} = \frac{U_{12}}{1+[\Pi_1+\Pi_2]U_{11}+ \Pi_1\Pi_2[U_{11}^2 - U_{12}^2]}, \notag  \\
U_{22}^{\rm scr} = \frac{U_{11} + \Pi_1[U_{11}^2 - U_{12}^2] }{U_{11} + \Pi_2[U_{11}^2 - U_{12}^2]} U_{11}^{\rm scr}.
\label{Ch1:DQW:ScrU22:ISB}
\end{gather}
In the diffusive regime ($ql\ll 1$, $\omega\tau_{\rm tr}\ll 1$) polarization operators become
\begin{equation}
\Pi_j(q,\omega) = \nu_\star \frac{D_j q^2}{D_j q^2-i\omega},\quad j=1,2 ,
\end{equation}
where $D_j$ denotes the diffusion coefficient in the $j$-th quantum well. We note that for $D_1\neq D_2$ dynamically screened inter well interactions in the left and right quantum wells are different. In the case of equal electron concentrations and mobilities $D_1=D_2=D$, $U_{11}^{\rm scr}=U_{22}^{\rm scr}$ and
\begin{gather}
U_{11/12}^{\rm scr}  = \frac{\varkappa (Dq^2-i\omega)}{2 \nu_\star q} \Biggl \{ \frac{1+e^{-qd}}{Dq\Bigl [q +\varkappa(1+e^{-qd})\Bigr]-i\omega} \notag\\
\pm 
\frac{1-e^{-qd}}{Dq\Bigl [q +\varkappa(1-e^{-qd})\Bigr]-i\omega} \Biggr \} .
\end{gather}
For $qd\gg 1$ electrons in the right quantum well do not affect inter well interaction in the left quantum well. In the opposite case, $qd\ll 1$, electrons in the right quantum well screen efficiently interaction between electrons in the left quantum well at $\varkappa d\lesssim 1$ only.

\subsubsection{Estimates for the interaction parameters \label{Ch1:DQW:Sec_Form_IntPar:ISB}}

\begin{figure}[t]
\centerline{\includegraphics[width=0.45 \textwidth]{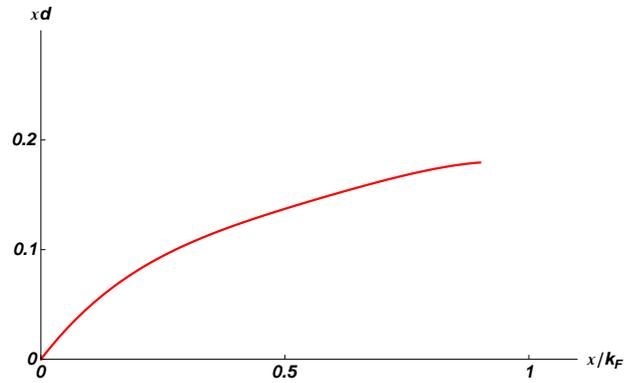}} 
\caption{Dependence of $\varkappa d_c$ on $\varkappa/k_F$. Adapted from Ref.\protect\cite{Burmistrov2011bISB} }
\label{Figure:Ch1:10:ISB}
\end{figure}

In the case of equal electron concentrations and mobilities  the interaction parameters $F_t$ and $F_v$ can be estimated with the help of Eqs.~\eqref{Ch1:DQW:ScrU22:ISB} as 
\begin{equation}
F_t\pm F_v = -\int\limits_0^{2\pi} \frac{d\theta}{4\pi} \frac{\varkappa (1\pm e^{-2k_Fd\sin\theta/2})}{2k_F\sin\frac{\theta}{2}+\varkappa  (1\pm e^{-2k_Fd\sin\theta/2})} . \label{Ch1:DQW:FtvPM:ISB}
\end{equation}
We note that RPA is justified provided $\varkappa/k_F\ll 1$. Equation~\eqref{Ch1:DQW:FtvPM:ISB} implies that $F_t$ and $F_v$ are negative and $|F_t| \geqslant |F_v|$. Interaction parameter $\tilde F_s$ changes sign from negative at small values of $d$ to positive at large values of $d$. The dependence of $d_c$ ($\tilde F_s(d_c)=0$) on $\varkappa/k_F$ is shown in Fig.~\ref{Figure:Ch1:10:ISB}. We mention that $|\tilde F_s|\leqslant |F_t|$ at $d<d_c$.

In the case of the single quantum well the interaction parameter in the triplet channel ($F_t^0$) is given by Eq.\eqref{Ch1:Ft0:ISB}. For quantum wells with equal electron concentrations and mobilities we find at $k_F d\gg 1$:
\begin{gather}
F_t = F_t^0 +\frac{1}{8\pi k_F d} \mathcal{G}_1(\varkappa d), \quad F_v = \frac{1}{8\pi k_F d} \mathcal{G}_2(\varkappa d) ,\notag \\ 
\mathcal{G}_1(x) = \frac{3 x\, e^x E_1(x)}{x+1} + \frac{2x\, e^{-2x/(x-1)}}{x^2-1} 
E_1\left (-\frac{2x}{x-1}\right ) , \notag \\
\mathcal{G}_2(x) = \mathcal{G}_1(x)-\frac{4 x\, e^x E_1(x)}{x+1} \label{Ftv:ISB} .
\end{gather}
Here $E_1(x)$ denotes the integral exponent, $E_1(x) = \int_x^\infty dt\, \exp(-t)/t$.

\subsection{One loop RG equations}

The matrix of interaction amplitudes $\mathbf{\Gamma}$ has the same structure as the matrix $\bf{F}$. It is useful to introduce the following notations ($a=1,2,3$ É $b=0,1,2,3$):
\begin{gather}
\Gamma_{10} = z \tilde\gamma_s= -\frac{z \tilde F_s}{1+\tilde F_s}, \quad \Gamma_{0a} = \Gamma_{1a}= z \gamma_t=-\frac{z F_t}{1+F_t}, \notag \\
 \Gamma_{2b} =\Gamma_{3b}=  z \gamma_v=-\frac{z F_v}{1+F_v} .\label{Ch1:DQW:DefG:ISB}
\end{gather}
We remind that $\Gamma_{00} = -z$. Due to the presence of finite spin ($\Delta_s$) and isospin ($\Delta_{SAS}$) splittings the NL$\sigma$M action \eqref{Ch1:Sstart:ISB} contains symmetry-breaking terms \eqref{Ch1:Sstartsb:ISB}. In what follows we consider temperatures $T \gg \Delta_{s,SAS}$ or, equivalently, length scales  $L\ll \sqrt{\sigma_{xx}/(16z \Delta_{s,SAS})}$ such that symmetry-breaking terms can be neglected.

Using Eqs. \eqref{Ch1:DQW:DefG:ISB} and \eqref{Ch1:RG_1loop:ISB} we find the following one loop RG equations for electrons in the symmetric double quantum well heterostructure :\cite{Burmistrov2011bISB}
\begin{align}
\frac{d \sigma_{xx}}{dy} &=- \frac{2}{\pi}\bigl [2+1+ f(\tilde{\gamma}_{s})+6 f(\gamma_t)+
8 f(\gamma_{v})  \bigr ] ,\label{Ch1:DQW:RG1_1:ISB}\\
\frac{d\tilde{\gamma}_s}{dy} &=
\frac{1+\tilde{\gamma}_s}{\pi\sigma_{xx}}\Bigl [ 
1-6\gamma_t-\tilde{\gamma}_s +8\gamma_v+
2 h(\tilde{\gamma}_s,\gamma_v)
\Bigr ] ,\label{Ch1:DQW:RG1_2:ISB}\\
\frac{d{\gamma}_t}{dy} &=
\frac{1+\gamma_t}{\pi\sigma_{xx}}\Bigl [1-\tilde{\gamma}_s +2\gamma_t +
h(\gamma_t,\gamma_v)
\Bigr ] ,
\label{Ch1:DQW:RG1_3:ISB}\\
\frac{d\gamma_v}{dy} &=
\frac{1}{\pi\sigma_{xx}} \Bigl [ (1+\tilde\gamma_s)(1-\gamma_v)+2\gamma_v p(\gamma_t,\gamma_v)\Bigr  ] ,
\label{Ch1:DQW:RG1_4:ISB}\\
\frac{d\ln z}{dy} &= \frac{1}{\pi\sigma_{xx}} \Bigl
[\tilde{\gamma}_s+6\gamma_t+8\gamma_v-1 \Bigr ] \label{Ch1:DQW:RG1_5:ISB}
\end{align}
where $h(x,y) = 8 y(x-y)/(1+y)$ and $p(x,y) = 1-3 x+4 y$. 
We emphasize that the r.h.s. of Eqs.~\eqref{Ch1:DQW:RG1_2:ISB} and \eqref{Ch1:DQW:RG1_3:ISB} is not polynomial in the intra well interaction amplitude $\gamma_v$. In all other known examples, the r.h.s. of one loop RG equations for the interaction amplitudes are second-order polynomials.\cite{FinkelsteinReviewISB, KirkpatrickBelitzISB, BurmistrovChtchelkatchev2008ISB,Punnoose2ISB,Punnoose3ISB}. This is intimately related with invariance of the NL$\sigma$M action $\mathcal{S}_\sigma+\mathcal{S}_F$ under transformations~\eqref{Ch1:GenFRot:ISB} of matrix $Q$. As follows from Eqs.~\eqref{Ch1:Falg:ISB}, the NL$\sigma$M action $\mathcal{S}_\sigma+\mathcal{S}_F$ with interaction amplitudes \eqref{Ch1:DQW:DefG:ISB} is invariant under rotations~\eqref{Ch1:GenFRot:ISB} with $\chi_{ab} = \chi \delta_{ac}\delta_{bd}$ where $c=0,1$, Á $d=1,2$ or $3$ for $\gamma_t=-1$. For $\tilde{\gamma}_s=-1$ the NL$\sigma$M action is invariant under rotations of $Q$ matrix with $\chi_{ab} = \chi \delta_{a1}\delta_{b0}$. This guaranties that $\gamma_t=-1$ and $\tilde{\gamma}_s=-1$ are fixed under the action of RG transformations. Therefore, the r.h.s. of RG equation for $\gamma_t$ ($\tilde{\gamma}_s$) should vanish at $\gamma_t=-1$ ($\tilde{\gamma}_s=-1$). For $\gamma_v=-1$ the NL$\sigma$M action $\mathcal{S}_\sigma+\mathcal{S}_F$ is not invariant under rotations~\eqref{Ch1:GenFRot:ISB}. Therefore, the r.h.s. of RG equations are not necessary finite at $\gamma_v=-1$.

Renormalization group Eqs.~\eqref{Ch1:DQW:RG1_1:ISB}-\eqref{Ch1:DQW:RG1_4:ISB} describes 4D ($\sigma_{xx}, \tilde\gamma_s, \gamma_t, \gamma_v$) flow diagram. The 2D surface $\gamma_t=\gamma_v=\tilde\gamma_s$ is invariant under RG flow. It corresponds to the case of coinciding quantum wells ($d=0$). In this case, RG Eqs.~\eqref{Ch1:DQW:RG1_1:ISB}-\eqref{Ch1:DQW:RG1_5:ISB} are equivalent to RG Eqs.~\eqref{Ch1:Si:RG0_1:ISB}-\eqref{Ch1:Si:RG0_3:ISB}. However, the surface $\gamma_t=\gamma_v=\tilde\gamma_s$ is unstable under perturbations in the initial values of interaction amplitudes due to finite value of $d$. There is the stable 2D surface $\gamma_v=0$, $\tilde\gamma_s=-1$ which is invariant under RG flow. It corresponds to the case of separate quantum wells ($d=\infty$). 

\begin{figure}[t]
\centerline{\includegraphics[width=0.5 \textwidth]{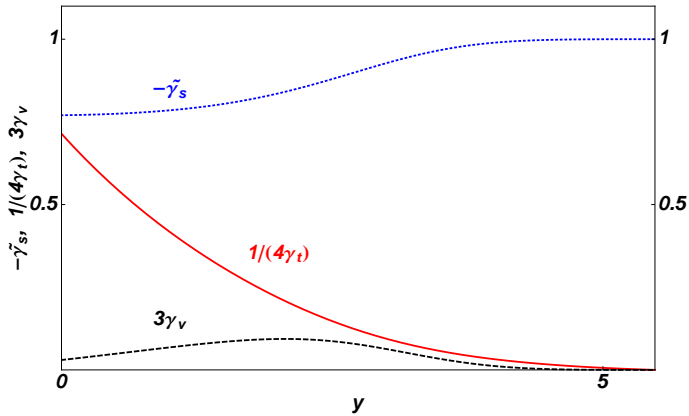}} 
\caption{Dependence of $\gamma_t$, $\gamma_v$, and $\tilde\gamma_s$ on $y$ for $\gamma_t(0)=0.35$, $\gamma_v(0)=0.01$, $\tilde\gamma_s(0)
=-0.77$, and $\sigma_{xx}(0)=6$.}
\label{Figure:Ch1:11:ISB}
\end{figure}

In addition, the 2D surface $\gamma_t=\tilde\gamma_s=-1$ and the line $\tilde\gamma_s = -1$, $\gamma_v=-1/2$, $\gamma_t=-1/3$ are invariant under RG flow in the one loop approximation. However, they are unreachable for the case of the double quantum well heterostructure since the following inequalities for the initial values of interaction amplitudes hold
$\gamma_t(0) \geqslant \gamma_v(0)\geqslant 0$, and $\gamma_t(0) \geqslant \tilde\gamma_s(0)$. Using Eqs.~\eqref{Ch1:DQW:RG1_2:ISB}-\eqref{Ch1:DQW:RG1_4:ISB}, one can check that conditions $\gamma_t \geqslant \gamma_v\geqslant 0$ and $\gamma_t \geqslant \tilde\gamma_s$ remains true under the RG flow. Interaction amplitude $\gamma_t$ increases always. Eventually, the renormalization group flows toward $\gamma_v=0$, $\tilde\gamma_s=-1$ and $\gamma_t=\infty$ as illustrated in Fig.~\ref{Figure:Ch1:11:ISB}. 

As follows from RG Eqs.~\eqref{Ch1:DQW:RG1_1:ISB}-\eqref{Ch1:DQW:RG1_4:ISB} the temperature dependence of conductivity  $\sigma_{xx}$ is of metallic type. Depending on the sign of the quantity $2+K_{ee}$ where  $K_{ee} =  1+f(\tilde \gamma_s(0))+6f(\gamma_t(0)) +8 f(\gamma_v(0))$, resistance either decreases monotonously under increase of $y$ (for $2+K_{ee}<0$) or has the maximum (for $2+K_{ee}>0$).


\subsection{Dephasing time}

The presence of electrons in the right quantum well changes the properties of electrons in the left quantum well.
Using Eq. \eqref{Ch1:DQW:TauPhi2:ISB} in the case of symmetric double quantum well with equal electron concentrations and mobilities we find the dephasing rate for electrons in one quantum well for $1/\tau_\phi \gg \Delta_{s,SAS}$\cite{Burmistrov2011bISB} 
\begin{gather}
\frac{1}{\tau_\phi} = \frac{\mathcal{A} T}{2\sigma_{xx}} \ln T \tau_\phi, \notag \\
\mathcal{A} = \frac{1}{2} \left [ 1 + \frac{\tilde{\gamma}_s^2}{2+\tilde{\gamma}_s}+6\frac{\gamma_t^2}{2+\gamma_t} +8 \frac{\gamma_v^2}{2+\gamma_v} \right ] .
\label{Ch1:DQW:TauPhi2:2:ISB}
\end{gather}
Here factor $1/2$ appears in the quantity $\mathcal{A}$ since Eq.\eqref{Ch1:DQW:TauPhi2:2:ISB} is the dephasing rate for electrons in one quantum well. Interaction amplitudes $\tilde \gamma_s$, $\gamma_t$, $\gamma_v$ and conductivity $\sigma_{xx}$ are taken at the length scale $L_T = \sqrt{\sigma_{xx}/z T}$. 

It is instructive to compare result~\eqref{Ch1:DQW:TauPhi2:2:ISB} with the result for the situation when one of the quantum wells is not filled by electrons. This corresponds to the limit $d\to \infty$. Using Eq.~\eqref{Ch1:DQW:TauPhi2:2:ISB} with $\tilde\gamma_s=-1$, $\gamma_v=0$, and $\gamma_t = \gamma_{t,0}$, we find~\cite{NarozhnyISB}
\begin{equation}
\mathcal{A} \,\to\, \mathcal{A}_0=  \left [ 1 + \frac{3\gamma_{t,0}^2}{2+\gamma_{t,0}} \right ],
\end{equation}
where initial value of $\gamma_{t,0}$ is equal to $\gamma_{t,0}(0)= -F_t^0/(1+F_t^0)$.


\subsection{Discussion and comparison with experiments}

Recently, weak localization and interaction corrections to conductivity have been studied experimentally\cite{MinkovGermanenko1ISB,MinkovGermanenko2ISB} in Al$_x$Ga$_{1-x}$As/GaAs/Al$_x$Ga$_{1-x}$As double quantum well heterostructure. The detailed investigation of two heterostructures, 3243 and 3154, with different dopping level has been performed. Analysis of experimental data on magnetoresistance allows to extract the dephasing time and the quantity  $K_{ee}$. The electron concentration in the right quantum well was controlled by a gate. In experiments of Refs.\cite{MinkovGermanenko1ISB,MinkovGermanenko2ISB} the electron concentration was high such that the total conductivity at high temperatures (about $4.2$) was about $80\, e^2/h$. As a consequence, interesting physics described by RG Eqs.~\eqref{Ch1:DQW:RG1_1:ISB}-\eqref{Ch1:DQW:RG1_5:ISB} should develop in the heterostructures 3243 and 3154 at very low temperature only and, therefore, was not observed in experiments of Refs.\cite{MinkovGermanenko1ISB,MinkovGermanenko2ISB}. The unexpected result found in Refs.\cite{MinkovGermanenko1ISB,MinkovGermanenko2ISB} is that the dephasing time (the coefficient $\mathcal{A}$ in Eq.~\eqref{Ch1:DQW:TauPhi2:ISB}) and interaction correction to the conductivity (the quantity $K_{ee}$) are almost independent of 
electron concentration in the right quantum well. The summary of experimental results of Refs.\cite{MinkovGermanenko1ISB,MinkovGermanenko2ISB} is presented in Table~\ref{Ch1:DQW:Tab1:ISB}. Theoretical estimates for the interaction parameters are summarized in Table~\ref{Ch1:DQW:Tab2:ISB}. 
In the experiments parameter $\varkappa d$ can be estimated as $\varkappa d=3.6$, interaction parameter $F_v$ is small, $F_t \approx F_t^0$, and $\tilde F_s\approx \varkappa d$. Comparison of theoretical and experimental estimates for quantities  $K_{ee}$, $\mathcal{A}$, $K_{ee,0}=1 +3 f(\gamma_{t,0}(0))$ and $\mathcal{A}_0$ are shown in Table~\ref{Ch1:DQW:Tab3:ISB}. š

\begin{table}[t]
\caption{Parameters for samples of Refs.\cite{MinkovGermanenko1ISB,MinkovGermanenko2ISB}. For both samples, the inverse screening length and distance between quantum wells are equal to $\varkappa=2.0\cdot 10^{6}$ cm$^{-1}$ and $d=1.8\cdot 10^{-6}$ cm, respectively.}
{\begin{tabular}{@{}ccccc@{}} \toprule
sample & $n$, cm$^{-2}$ &$k_F$,  cm$^{-1}$ &   $k_F d$ & $\varkappa/k_F$\\
\colrule
\#3154 & $4.5\cdot 10^{11}$ & $1.7\cdot 10^{6}$ &   $3.06$ & $1.18$ \\
\#3243 &  $7.5\cdot 10^{11}$ & $2.2\cdot 10^{6}$ &  $3.95$  & $0.91$ \\
\botrule
\end{tabular}}
\label{Ch1:DQW:Tab1:ISB}
\end{table}

\begin{table}[t]
\caption{Theoretical estimates for interaction amplitudes in samples  3154 and 3243 of Refs.\cite{MinkovGermanenko1ISB,MinkovGermanenko2ISB}.}
{\begin{tabular}{@{}ccccc@{}} \toprule
sample &  $\tilde \gamma_s(0)$ & $\gamma_t(0)$ & $\gamma_v(0)$ & $\gamma_{t,0}(0)$ \\ 
\colrule
\#3154 &  $-0.77$ & $0.35$ & $0.009$ &  $0.35$ 
\\
\#3243  & $-0.77$ & $0.30$ & $0.007$ &  $0.30$ \\
\botrule
\end{tabular}}
\label{Ch1:DQW:Tab2:ISB}
\end{table}


\begin{table}[t]
\caption{Comparison theoretical and experimental estimates for quantities $K_{ee}$, $\mathcal{A}$, $K_{ee,0}$ and $\mathcal{A}_0$.}
{\begin{tabular}{@{}ccccc@{}} \toprule
& \multicolumn{2}{c}{Theory} & \multicolumn{2}{c}{Experiment}  \\
& \#3154 & \#3243& \#3154 & \#3243 \\
\colrule
$K_{ee}$ & $0.59$ & $0.72$ &$0.50\pm 0.05$& $0.57\pm 0.05$ \\
$K_{ee,0}$ &$0.52$& $0.59$ &$0.53\pm 0.05$& $0.60 \pm 0.05$ \\
$\mathcal{A}$ & $0.89$ & $0.86$ && \\
$\mathcal{A}_0$ &$1.15$ & $1.12$ && \\ 
$\mathcal{A}/\mathcal{A}_0$ & 0.77 & 0.77 & $1.00\pm 0.05$ & $1.00\pm 0.05$ \\
\botrule
\end{tabular}}
\label{Ch1:DQW:Tab3:ISB}
\end{table}

As one can see from the Table~\ref{Ch1:DQW:Tab3:ISB}, theoretical estimates are in good quantitative agreement with the experimental findings. We note that since $K_{ee,0}>0$ for $\varkappa/k_F\lesssim 1$, the interesting situation can be realized in the double quantum well heterostructure with $\varkappa d \lesssim 1$ in which in the case of equal electron concentration and mobilities one expects  $K_{ee} < 0$.

The theory developed above is valid at temperatures $T\gg \Delta_{SAS}, \Delta_s, 1/\tau_{+-}$. In experiments of Refs.\cite{MinkovGermanenko1ISB,MinkovGermanenko2ISB} the spin splitting (for magnetic fields in which
magnetoresistance has been measured) and the symmetric-antisymmetric splitting can be estimated as  $\Delta_s \lesssim 0.2$ K and $\Delta_{SAS} \lesssim 1$ K. Small asymmetry in impurity distribution along $z$ axis with respect to the middle plain between quantum wells yields elastic scattering between symmetric and antisymmetric states. Corresponding rate ($1/\tau_{+-}$) can be estimated from experimental data of Refs.\cite{MinkovGermanenko1ISB,MinkovGermanenko2ISB} on magnetoresistance. As well-known,
\cite{Aronov-AltshulerISB} in the absence of scattering between symmetric and antisymmetric states ($1/\tau_{+-}=0$), the presence of $\Delta_s$ and/or $\Delta_{SAS}$ does not influence weak localization correction. In the absence of magnetic field the weak localization correction in both limiting cases,
 $\Delta_{SAS}\ll 1/\tau_{+-}$ and $\Delta_{SAS}\gg 1/\tau_{+-}$ can be written as\cite{Burmistrov2011bISB} 
\begin{gather}
\delta\sigma_{xx}^{WL} = \frac{1}{\pi} \ln \Bigl [ \frac{\tau_{\rm tr}^2}{\tau_\phi}\Bigl (\frac{1}{\tau_\phi}+\frac{1}{\tau_{12}}\Bigr )\Bigr ] , \notag \\
 \frac{1}{\tau_{12}} \sim \min \left \{\Delta_{SAS}^2\tau_{+-},\tau^{-1}_{+-}\right \} .\label{Ch1:DQW:WL1:ISB}
\end{gather}
Equation~\eqref{Ch1:DQW:WL1:ISB} interpolates between the result for two valleys at high temperatures ($1/\tau_\phi\gg 1/\tau_{12}$) and for single valley at low temperatures ($1/\tau_\phi\ll 1/\tau_{12}$). For experiments of Ref.\cite{MinkovGermanenko1ISB} the rate $1/\tau_{12}$ can be estimated as about $0.1$ K. Using the estimate $\Delta_{SAS} \lesssim 1$ K, we find that $1/\tau_{+-} \sim 1/\tau_{12}\lesssim 0.1$ K. Therefore, the theory described above is applicable to the experiments of Refs.\cite{MinkovGermanenko1ISB,MinkovGermanenko2ISB} at temperatures $T\gtrsim 1$ K. This is the range in which experiments of Refs.\cite{MinkovGermanenko1ISB,MinkovGermanenko2ISB} have been performed.

\section{Conclusions \label{Conc.Ch1:ISB}}

Based on the NL$\sigma$M approach and the renormalization group treatment we study the effect of spin and isospin degrees of freedom on low temperature transport in 2D strongly interacting disordered electron system. In this case we derive general renormalization group equations in the one loop approximation. We find that the standard case  of equal interaction amplitudes for interaction among electrons with different spin and isospin projections is  
unstable. We demonstrate that such situation (with different interaction amplitudes) can be naturally realized in 2D two-valley electron system in Si-MOSFET and in 2D electron system in double quantum well heterostructure with common disorder. Using general one loop RG equations we explain experimentally observed in Si-MOSFET variation of temperature dependence of resistivity from metallic to insulating with increase of parallel magnetic field. For Si-MOSFET we predict the temperature dependence of resistivity with two maximums. 
In double quantum well heterostructure the general one loop RG equations allow us to predict that in spite of common scatters electrons in each quantum well become independent  at low temperatures.

\section*{Acknowledgements}

I am grateful to my coauthors N.M. Chtchelkatchev, I.V. Gornyi, D.A. Knyazev, O.E.  Omel'yanovskii, V.M. Pudalov, K.S. Tikhonov for fruitful collaboration on the subjects discussed in this review. I am indebted to A.V. Germanenko, D.A. Knyazev, A.A. Kuntsevich, G.M. Minkov,  O.E.  Omel'yanovskii, V.M. Pudalov, and A.A. Sherstobitov for detailed discussions of their experimental data prior to publication. I thank A.S. Ioselevich, A.M. Finkelstein, A.D. Mirlin, P.M. Ostrovsky, A.M.M. Pruisken, M.A. Skvortsov, and A.G. Yashenkin for useful discussions and comments. The work was partially supported by Russian President Grant No. MD-5620.2016.2, Russian President Scientific Schools Grant NSh-10129.2016.2, and RFBR Grant No. 15-32-20176.

\appendix
\section{Appendix \label{App:Ch1:ISB}}

In this section we analyze the stability of the fixed point with $\Gamma_{ab}=\Gamma$ for $(ab)\neq (00)$ of
the one loop RG Eqs. \eqref{Ch1:RG_1loop:ISB}.  Let us write $\gamma_{ab}=\gamma+\eta_{ab}$ for $(ab)\neq (00)$ and $\gamma_{00}=\gamma_{00}+\eta_{00}$ where quantities $\eta_{ab}$ are assumed to be small. Linearizing Eqs. \eqref{Ch1:RG_1loop:ISB} we find 
\begin{align}
\frac{d \eta_{ab}}{dy}  & = \frac{1}{\pi\sigma_{xx}} \Biggl [ (17\gamma-\gamma_{00})\eta_{ab} 
- \sum_{cd} \eta_{cd}    \bigl ( \gamma
\notag \\ 
 & +\frac{1}{4} \Sp [t_{ab}t_{cd}]^2\bigr )\Biggr ], \qquad (ab)\neq (00)
\end{align}
and 
\begin{equation}
\frac{d\eta_{00}}{dy}  = -\frac{1}{\pi\sigma_{xx}} \Biggl [ (15\gamma+\gamma_{00})\eta_{00}+(1+\gamma_{00})\sum_{cd}\eta_{cd}
\Biggr ] .
\end{equation}
It is convenient to introduce the following variables $\mu_{ab} =\sum_{cd\neq(00)} \Sp [t_{ab}t_{cd}]^2 \eta_{cd}/4$ and $\mu_{00}=\sum_{cd\neq(00)} \eta_{cd}$. We mention that $\mu_{00}$ is projection of the 16-dimensional vector $\{\eta_{ab}\}$ on the direction $\mathbf{e}=\{0,1,\dots,1\}$. Then for $(ab)\neq (00)$ we obtain
\begin{equation}
\frac{d}{dy} \begin{pmatrix}
\eta_{ab}\\
\mu_{ab} \\
\eta_{00} \\
\mu_{00}
\end{pmatrix} = \frac{1}{\pi\sigma_{xx}} 
\begin{pmatrix}
\delta_1 & -1 & -\beta & -\gamma \\
-16 & \delta_1 & \beta & \beta \\
0& 0 & -\delta_2 & -\beta_{00} \\
0 &0& -15\beta & \delta_3
\end{pmatrix} 
\begin{pmatrix}
\eta_{ab}\\
\mu_{ab} \\
\eta_{00} \\
\mu_{00}
\end{pmatrix}  \label{App:Ch1:RG_Analysis:M:ISB}
\end{equation}
where $\beta=1+\gamma$, $\beta_{00}=1+\gamma_{00}$, $\delta_1= 17\gamma-\gamma_{00}$, $\delta_2=15\gamma+2\gamma_{00}+1$ and $\delta_3=2\gamma-\gamma_{00}+1$.
The behavior of quantities $\eta_{ab}$  with $(ab)\neq (00)$ in the direction perpendicular to the plane based on vectors $\mathbf{e}$ and $\mathbf{e}_0=\{1,0,\dots,0\}$ is characterized by eigenvalues  $\lambda_{\pm} = 17\gamma-\gamma_{00}\pm 4$. They are positive at large enough values of  $\gamma$. This means that in Eqs. \eqref{App:Ch1:RG_Analysis:M:ISB} small quantities  $\eta_{ab}$ satisfying $\eta_{00}=\mu_{00}=0$ increase as the length scale $y$ grows. Therefore the fixed point with $\Gamma_{ab}=\Gamma$ for $(ab)\neq (00)$ of
the one loop RG Eqs. \eqref{Ch1:RG_1loop:ISB} is unstable.


\end{document}